\documentclass[11pt]{article}

\usepackage{fullpage}
\usepackage[pdftex]{graphicx}
\usepackage{url, float, lscape, verbatim, bigints}
\usepackage{xcolor, dsfont, pdfpages, sectsty, MnSymbol}
\usepackage{bbm, dsfont}
\usepackage{amsmath}
\usepackage{psfrag,epsf}
\usepackage{latexsym}
\usepackage{calc}
\usepackage[ragged, symbol]{footmisc}
\usepackage{grffile}
\usepackage{booktabs}
\usepackage{algorithmic}
\usepackage{algorithm}
\usepackage{caption}
\usepackage{comment}
\usepackage{color}

\usepackage{breakcites}
\usepackage{etoolbox}
\makeatletter
\patchcmd{\@citex}{,}{;}{}{}
\makeatother

\newcommand\pN{\mathcal{N}}
\newcommand{\blind}{0}

\usepackage[left=1.1in, right = 1.1in, top = 1.1in, bottom = 1.1in]{geometry}
\renewcommand{\baselinestretch}{1.3}

\begin{document}

	\def\spacingset#1{\renewcommand{\baselinestretch}%
		{#1}\small\normalsize} \spacingset{1}


	\if0\blind
	{
		\title{\bf Classification of pediatric pneumonia using chest X-rays by functional regression}
		\author{Md Nazmul Islam\thanks{
				UnitedHealth Group R\&D (Email: \textit{nazmulislam@uhg.com})}
}			
  \date{}
  \maketitle
} \fi
	
	\if1\blind
	{
		\bigskip
		\bigskip
		\bigskip

\begin{center}
	{\bf \Large  Classification of pediatric pneumonia using chest X-rays by functional regression}\\ 
\end{center}
 \medskip
} \fi

\baselineskip=16pt

\section*{Abstract}
An accurate and prompt diagnosis of pediatric pneumonia is imperative for successful treatment intervention. One approach to diagnose pneumonia cases is using radiographic data. In this article, we propose a novel parsimonious scalar-on-image classification model adopting the ideas of functional data analysis. Our main idea is to treat images as functional measurements and exploit underlying covariance structures to select basis functions; these bases are then used in approximating both image profiles and corresponding regression coefficient. We re-express the regression model into a standard generalized linear model where the functional principal component scores are treated as covariates. We apply the method to (1) classify pneumonia against healthy and viral against bacterial pneumonia patients, and (2) test the null effect about the association between images and responses. Extensive simulation studies show excellent numerical performance in terms of classification, hypothesis testing, and efficient computation.

\textbf{\underline{Keywords:}} X-rays; Image classification; Functional data analysis; Pneumonia detection; Scalar-on-image regression.

\section{Introduction}\label{s:Intro}

Pneumonia is a form of an acute respiratory illness affecting the lungs. Though pneumonia can occur at any age, younger children are most vulnerable. Bacteria, viruses, or fungi can categorize the common causes of pneumonia. In all these cases, children experience breathing difficulty as their lungs get contaminated with pus and fluid, which results in the inflammation of air sacs limiting the oxygen intake. Pediatric pneumonia is the primary cause of fatality that claims 3 million children globally every year, according to the United Nations Children's Fund (UNICEF) \cite{unicef2019}. Unfortunately, the majority of these deaths were preventable and occurred due to the lack of proper immunization, adequate nutrition, and the availability of antibiotic treatments. The developing countries, particularly in Southeast Asia and sub-Saharan Africa, suffer most and have the majority of death tolls. The developed countries still suffer substantially from the burden of disease, with 2.5 million incidences yearly leading a third to a half of these to hospitalizations incurring high healthcare-associated costs; see \cite{bennett2011pediatric, howie2019global, ebeledike2019pediatric}. Very recently, with the discovery of the pneumococcal vaccine, the risk of pneumonia in the United States has been reduced substantially. In contrast, the developed countries are yet to avail such measures, and they have been working strenuously to achieve the Integrated Global Action Plan for the Prevention and Control of Pneumonia and Diarrhoea (GAPPD) under the supervision of the World Health Organization (WHO) and UNICEF.

The clinical signs or symptoms for pneumonia are often nonspecific and vary based on the patient's physical characteristics. Accurate and timely diagnosis is the key to fight pneumonia. The current state-of-the-art diagnostic approaches include polymerase chain reaction (PCR), serology, culture, complete blood cell (CBC), chest radiography, and ultrasonography based test. Some of these tests require advanced lab facilities and are not entirely immune from producing false negatives. The PCR test is widely used but requires 24-48 hours to get back the diagnostic results and often requires close contact to collect the specimens from the suspected individuals manually, which is risky and has a high chance of transmitting infectious diseases like COVID-19. Recently, chest X-rays have shown a lot of promise and been used in diagnosis or as a confirmatory test for pneumonia \cite{markowitz1998pneumonia, kim2007round}. Such medical images need to be interpreted by radiologists to detect appropriately any abnormalities; unfortunately, this is often a manual, time-consuming step, and requires experts with domain-knowledge to diagnose. Therefore, there is a need for intelligent decision support that can empower and augment clinical decision making, which in turn saves time and prevents physicians' burnout and morbidity. Such a tool, if developed, may classify, flag, or confirm pneumonia-like patients using radiography images or can even stratify symptomatic patients into different risk categories as needed; see \cite{bar2015deep, kermany2018identifying, wang2018tienet, terzopoulos2019semi, wang2020covid,gozes2020coronavirus} and many others. 


Deep learning-based artificial intelligence (AI) systems have been used extensively in studying radiology images and detecting patients \cite{kermany2018identifying}. While AI works well in prediction, it requires a substantially large dataset to train the machine learning model, which may not always be feasible. Besides, such an approach does not offer flexibility to either make inference about the parameters of interest or quantify uncertainty in estimation. Very recently, statistical-based methodologies have been applied in neuroimaging; see \cite{shin2010voxel, kang2011meta, huang2013bayesian, wang2014regularized, reiss2015wavelet} and many others. \cite{penny2005bayesian, kang2011meta, huang2013bayesian} adopted Bayesian spatial modeling approaches to model the correlation between neighboring voxels with carefully placing priors on brain regions in predicting adverse clinical outcomes of Alzheimer's disease. In a similar spirit, \cite{smith2006tract,grimmer2009clinical,shin2010voxel} fitted univariate linear regressions on each region of interest of brain image separately to explore the relationship between images and scalar measures. While most of these methods primarily focus on prediction, the inferential aspects of parameter estimates are not discussed explicitly. \cite{wang2014regularized, wang2017classification} model an image data as functional measurements and apply the regularization penalty to induce zero-effect for the region where there is no association between responses and images. Here the authors used pre-specified Haar wavelet basis functions to approximate coefficient functions for which the selection of the optimum number of basis is based on either information criterion or cross-validation, which is a time-consuming step for multi-dimensional images. Such methodologies are also not directly applicable in the presence of noisy image data. Furthermore, the authors adopted a permutation-based test to make inference about the non-zero effect in a region, and it is unclear how to test the nullity of the overall regression coefficient.

In this paper, we address these limitations and introduce a novel scalar-on-image regression procedure based on functional principal component (FPC) analysis. There are three key novelties of this paper. The first novelty is the use of the functional data analysis approach exploiting the underlying covariance structure of 2-dimensional (2D) X-ray data, which has not previously been used, to the best of our knowledge, in this generality in image classification. The second novelty is the parsimonious modeling of the coefficient function using data-driven basis functions. The third novelty is casting the scalar-on-image regression model into a generalized linear model framework, which allows hypothesis testing about the parameters of interests explaining the association between responses and covariates. The main advantages of this approach are as below - (1) It provides excellent numerical performance in terms of classification accuracy; (2) it is computationally efficient and magnitude faster; (3) unlike the deep learning approach, it does not require a massive set of images to train the model; (4) it offers inference about the effect of radiographic images in classification; (5) it is applicable for images corrupted with subtle noises.

The paper is organized as follows. Section \ref{s:data} describes the collection and processing of the X-ray. Section \ref{s:propose} details the image classification approach. Section \ref{s:estimatebasis} presents the estimation procedure. Section \ref{s:inference} describes the inferential procedure. Section \ref{s:data analysis} presents the application in classifying pediatric chest X-rays to detect pneumonia. Multiple simulation studies are performed in Section \ref{s:simulation}, showing the robustness of our method mimicking the original data. Section \ref{s:discuss} concludes the paper with a discussion.

\section{Data processing}\label{s:data}

\begin{figure}[ht]
	\centering
	\caption{Illustrations of chest X-ray images for four randomly chosen healthy subjects (top panel) and four pneumonia (bottom panel) patients, where the two bottom-left panels associate with the bacterial pneumonia cases, and the two bottom-right refers to the viral pneumonia cases. Note that all patients are aged between 1-5 years.}
	\includegraphics[width=0.22\textwidth]{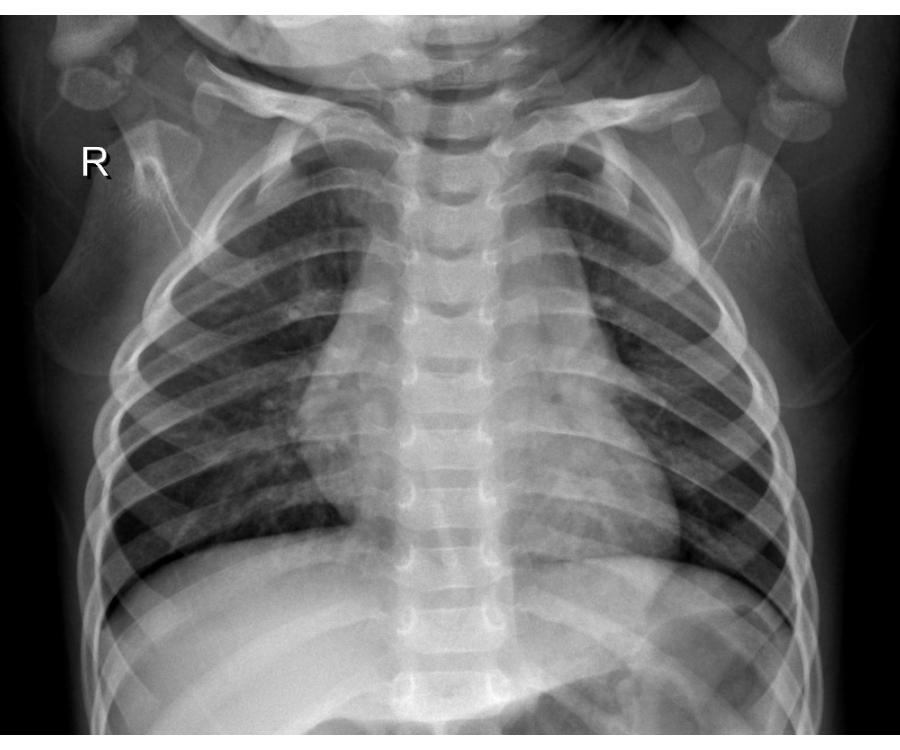}
	\includegraphics[width=0.22\textwidth]{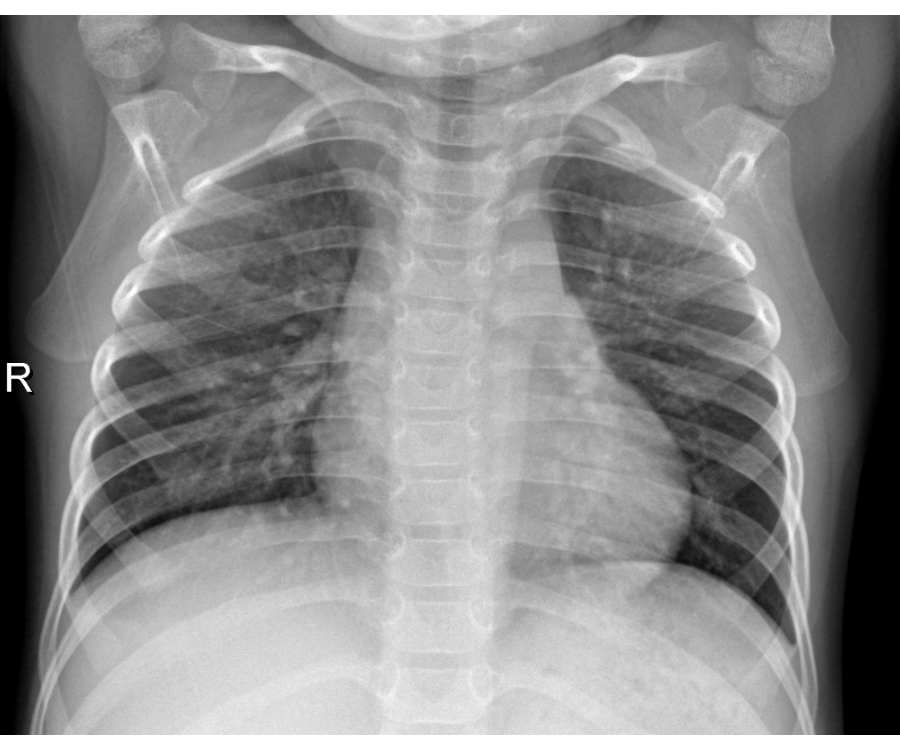}
	\includegraphics[width=0.22\textwidth]{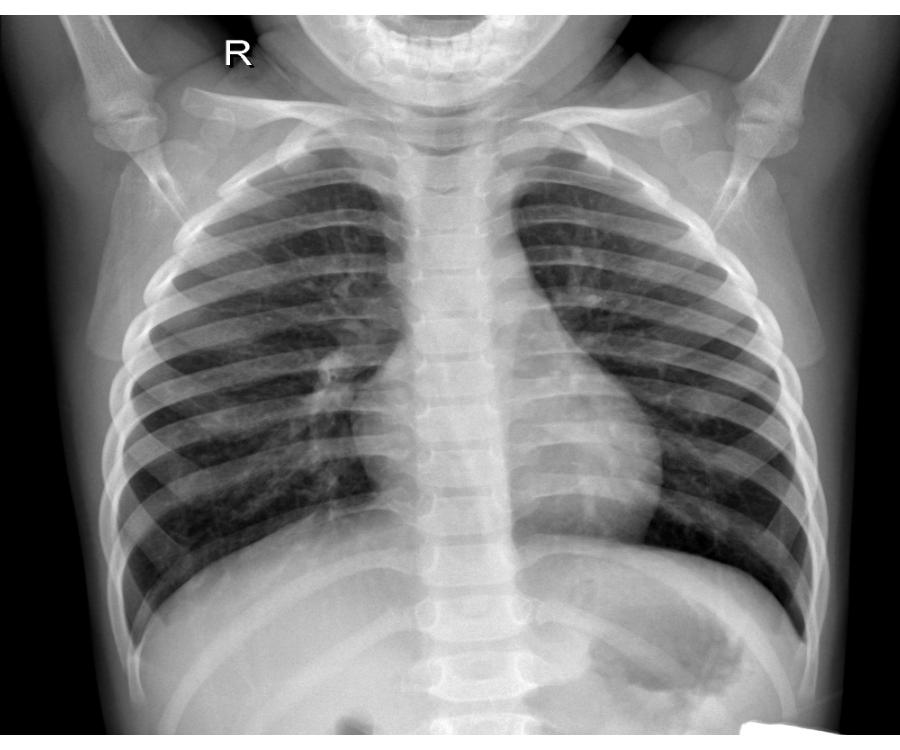}
      \includegraphics[width=0.22\textwidth]{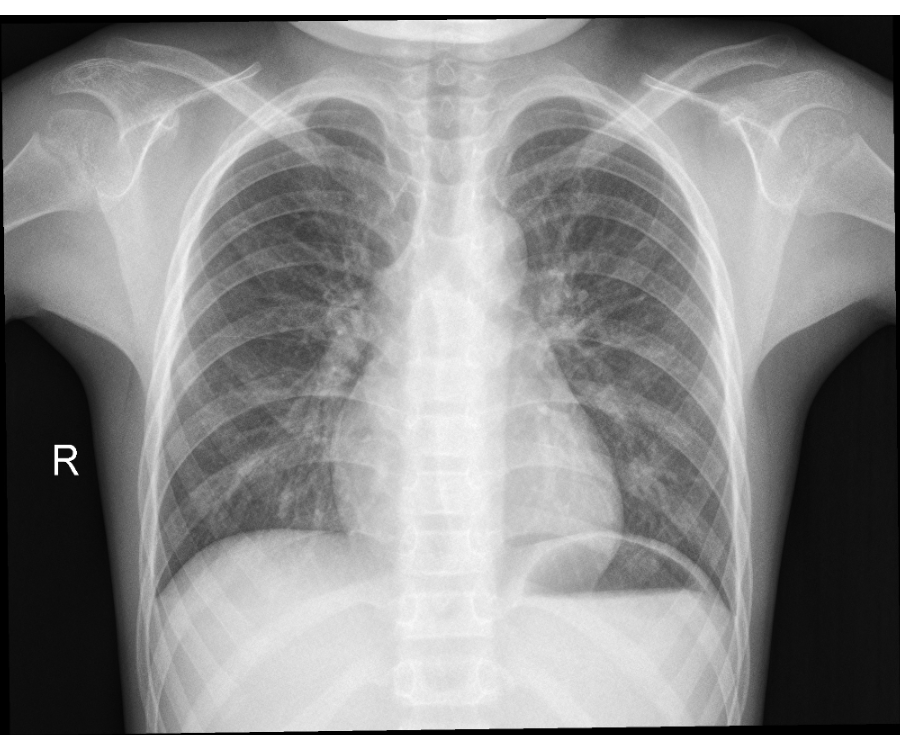}\\
	\includegraphics[width=0.22\textwidth]{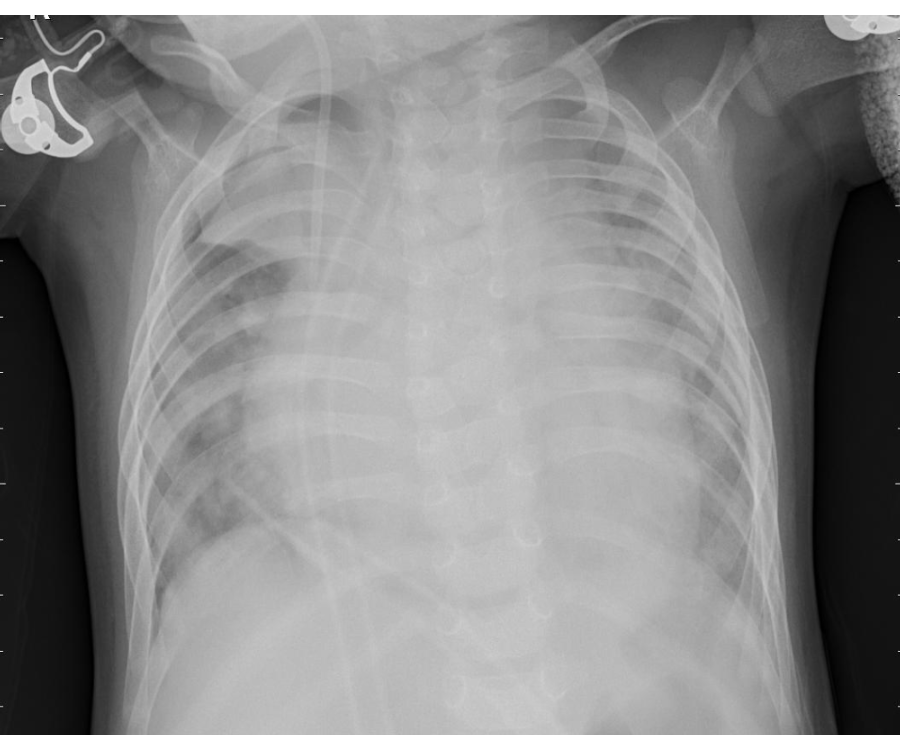}
       \includegraphics[width=0.22\textwidth]{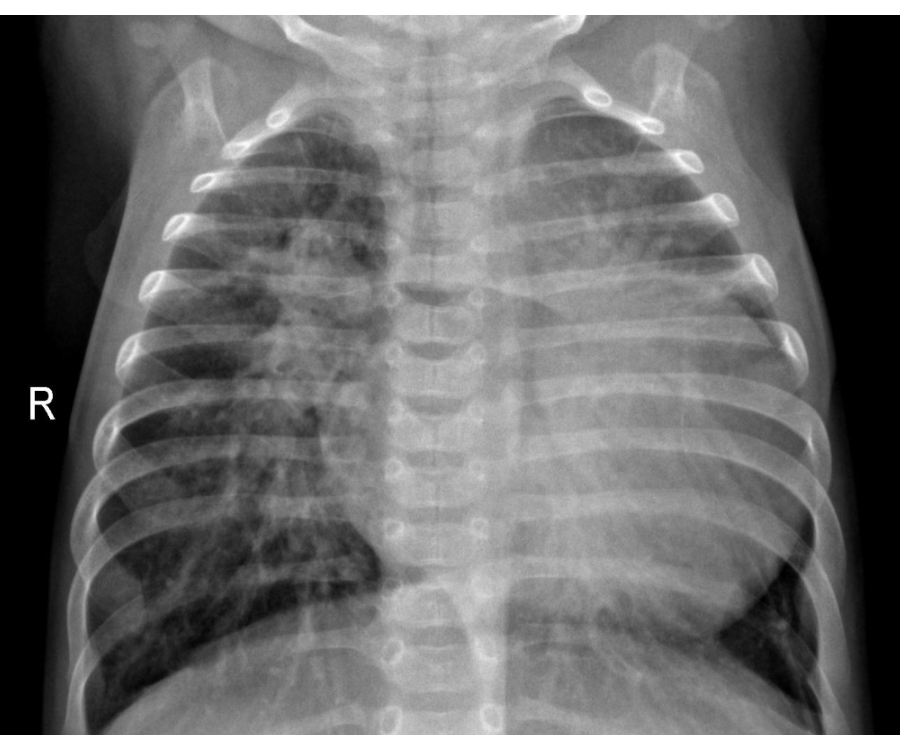}
	\includegraphics[width=0.22\textwidth]{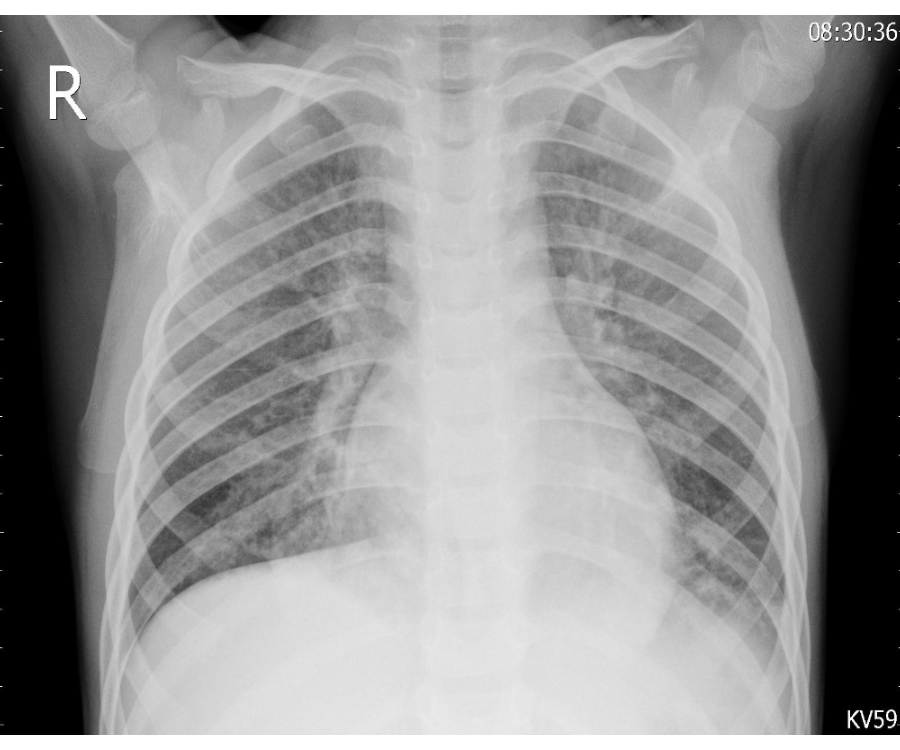}
	\includegraphics[width=0.22\textwidth]{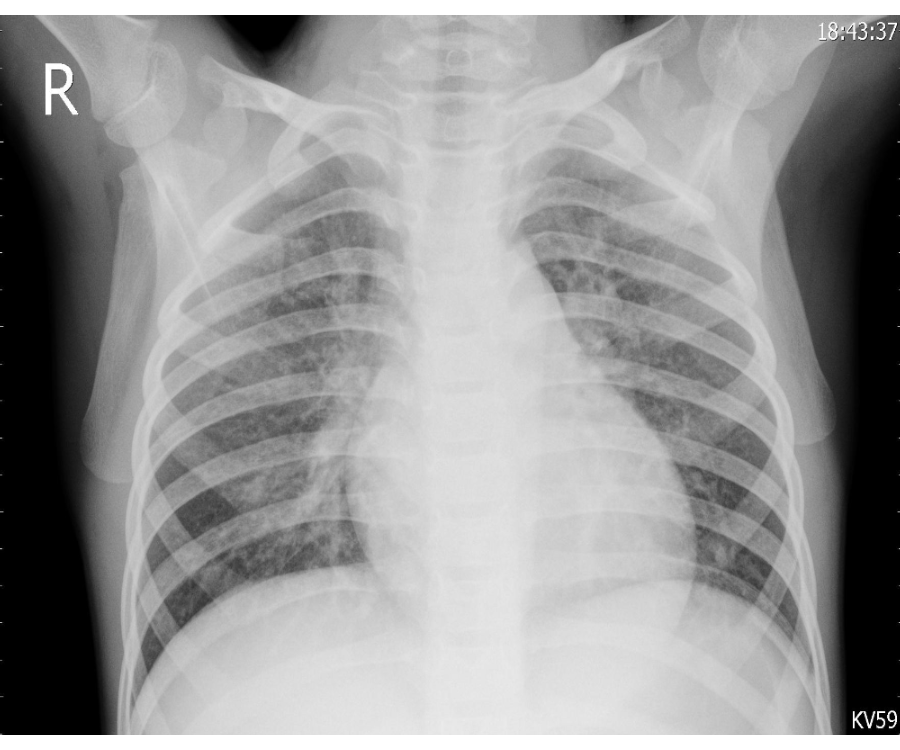}
	\label{f:xray} 
\end{figure}

Figure \ref{f:xray} illustrates the posteroanterior X-ray images for (a) four randomly chosen healthy subjects (top four panels), (b) two bacterial pneumonia patients (two-bottom-left), and (c) two viral pneumonia patients (two-bottom-right). (a) The normal chest X-rays have no signs of abnormal opacification such as fluid or solid materials within the airways in the images, (b) Bacterial pneumonia usually displays a focal lobar consolidation where a segment of lung shows confluent cotton-like opacity, (c) Apparently, more diffuse ``interstitial" patterns such as scattered white patches in both lungs are evident in the X-rays of viral pneumonia patients.

The pneumonia dataset \cite{datapneumonia} consists of 5,863 X-Ray images (in JPEG format) for two categories: pneumonia (bacterial and viral) and normal cases; see Figure \ref{f:xray}. This dataset is freely available in {\tt Kaggle}. These chest X-ray images (anterior-posterior) were collected from pediatric patients aged between one to five years who were treated in the Guangzhou Women and Children's Medical Center, Guangzhou, China, before 2018. These chest X-ray images were recorded during the patients' routine clinical visits. For the purpose of analysis, all chest radiographs were initially screened for quality control by removing all low quality or unreadable scans, and the diagnoses were labeled by two expert physicians \cite{datapneumonia}. For the purpose of our analysis, we use the X-ray images saved in the \textit{Train} folder as there could be an issue with labeling in the \textit{Test} folder identified by independent researchers; see \cite{datapneumonia_discussion}. There are 5,216 images (1,341 normal and 3,875 pneumonia cases) in the folder, which we use for our analysis, assuming that the current labeling is correct.

Each X-ray image is grayscaled 2-dimensional (2D) array structure of dimension $U_{i} \times V_{i}$ where $U_{i}$ and $V_{i} $ represents the width and height of the images. The image data are comprised of intensity profiles with values between 0 and 1 representing black and white, respectively. We reshape all images into the same size, i.e., $120 \times 80$ for our computational convenience.

\section{Proposed methodology}\label{s:propose} 
Let the observed data be $\big[y_{i}, \{Z_{ir}; r = 1, \ldots, R\}, \{X_{i}(h_{u},w_{v}); u = 1, \ldots, U, v = 1, \ldots, V\}; i = 1, \ldots, N\big];$ where $i$ indexes the subject, $y_{i}$ is the scalar response associated with the $i$th subject, $Z_{ir}$ is the $r$th observed clinical or demographic (e.g, age, prior history of lung diseases, number of days since the start of symptoms or hospitalization),  $X_{i}(h_{u}, w_{v})$ is the realized discrete intensity value of a two-dimensional (2D) projection image at discrete grid points $h_{u}$ and $w_{v}$ such that $h_{u} \in \mathcal{H}$ and $w_{v} \in \mathcal{W};$ where both $\mathcal{H}$ and $\mathcal{W}$ are closed compact sets. We assume that $\{h_{1}, \cdots, h_{U}\}$  and $\{w_{1}, \cdots, w_{V}\}$ are regular, dense design in $\mathcal{H}$ and  $\mathcal{W},$ respectively. In our application, $y_{i}$ takes values either 0 or 1 indicating the disease status (i.e., normal or pneumonia) of the $i$th subject and $X_{i}(\cdot, \cdot)$'s indicate the intensity values of the X-ray images which are the 2D projections of 3D objects. In our modeling, we treat each 2D image object as a functional measurement. 

Denote by $EF\{\mu_{i}, \eta\}$ the exponential family distribution with mean $\mu_{i}$ and dispersion parameter $\eta.$ As our objective is to classify patients with respect to their radiograph images, we propose to fit the following generalized functional regression model 
\begin{equation}
\begin{aligned}
\label{e:FLM}
y_{i}|X_{i}, Z_{i} &\sim EF(\mu_{i}, \eta),\\
g(\mu_i) &= \beta_{0} + \sum^{R}_{r=1}Z_{ir}\beta_{r}  +  \int_\mathcal{H} \int_\mathcal{W} X_{i}(h, w) \gamma(h, w) dw dh;
\end{aligned}
\end{equation}
where $g(\cdot)$ is a known monotone link function (e.g., logit link for the binary responses), $\beta_{0}$ is an unknown scalar intercept, and $\gamma(\cdot, \cdot)$ is an unknown bivariate function defined on $\mathcal{H} \times \mathcal{W}$ quantifying the association between mean responses and functional covariate. If the chest X-rays of healthy and pneumonia afflicted patients provide non-distinguishable pattern, then $\gamma(\cdot, \cdot) = 0.$ For notational convenience, we denote $\gamma(\cdot, \cdot)$ simply by $\gamma$ and use them interchangeably. Model \ref{e:FLM} is a variation of the classical functional linear models described in \cite{cardot1999functional,  james2005functional, reiss2007functional, goldsmith2011penalized, muller2012functional, gertheiss2013longitudinal} and also an extension of the functional varying coefficient models \cite{cardot2008varying, kundu2016longitudinal, staicu2020longitudinal, stalling2020}.     


\subsection{Model approximation}\label{ss:modelapprox}
Let $\{\phi_{k}(\cdot): k \geq 1\}$ be a set of orthonormal basis defined in $L^{2}(\mathcal{W})$ such that $\int_{\mathcal{W}} \phi_{k}(w) \phi_{k'}(w) dw = 0$ if $k \neq k'$ and 1 otherwise. Similarly, define by $\{\psi_{l}(\cdot): l \geq 1\}$ the orthonormal basis functions spanned in $L^{2}(\mathcal{H})$ such that  $\int_{\mathcal{H}} \psi_{l}(h) \psi_{l'}(h) dh = 0$ if $l \neq l'$ and 1 otherwise. We use the tensor product of these basis functions to approximate $\gamma$ such that $\gamma(s,t) = \sum_{k} \sum_{l} \phi_{k}(w)  \psi_{l}(h) \beta_{kl},$ where $\beta_{kl}$'s are the unknown basis coefficients and can be defined uniquely by $\beta_{kl} = \int_{\mathcal{H}} \int_{\mathcal{W}} \gamma(s,t) \phi_{k}(w) \psi_{l}(h)  dw dh.$ For our modeling exposition, we let $\psi_{l}(\cdot)$ vary within each $k$ resulting in $\gamma(s,t) = \sum_{k} \sum_{l} \phi_{k}(w)  \psi_{kl}(h) \beta_{kl}.$ We represent the functional covariate using the same basis functions $\{\phi_{k}(\cdot)\}_{k\geq 1}$ and $\{\psi_{kl}(\cdot)\}_{l\geq 1}$ in two steps: first, we write $X_{i}(h, w) = \sum_{k} \xi_{ik}(h) \phi_{k}(w)$ where $\xi_{ik}(\cdot)$ is a zero-mean smooth process defined on $\mathcal{H},$ and independent identically distributed (IID) over $i.$ Next, for each $k,$ write $\xi_{ik}(h) = \sum_{l} \zeta_{ikl} \psi_{kl}(h),$ where $\zeta_{ikl}$'s are the weights associated with the corresponding basis functions $\psi_{kl}(\cdot)$ and IID over $i$ and $l$. It follows that 
    \begin{equation}
    \begin{aligned}
    \label{e:FLM_approx}
    & \int_\mathcal{H} \int_\mathcal{W} X_{i}(h, w) \gamma(h, w) dw dh \\ =& \int_\mathcal{H} \int_\mathcal{W} \Big[\sum_{k \geq 1}   \sum_{l \geq 1} \zeta_{ikl} \psi_{kl}(h)  \phi_{k}(w) \Big] \Big[ \sum_{k \geq 1} \sum_{l \geq 1} \phi_{k}(w)  \psi_{kl}(h) \beta_{kl} \Big] dw dh, \\
    =& \sum_{k \geq 1}   \sum_{l \geq 1} \zeta_{ikl} \beta_{kl}.
    \end{aligned}
    \end{equation}
Here the infinite summation is intractable and needs to be truncated at some finite levels, say $K$ and $L$ for $w$-direction and $h$-direction, respectively. In a similar spirit to \cite{staicu2020longitudinal}, the truncation values $K$ and $L$ control the degree of smoothness of $\gamma$ in both directions. As in the non-parametric regression approach, the choice of truncation values leads to overly smooth or wiggly patterns in $\gamma$ and therefore needs to be selected with caution that we discuss in the next section. Using the equation \ref{e:FLM_approx}, we cast the model into a generalized linear model \cite{nelder1972generalized} framework for given $\zeta_{ikl}$'s as below
\begin{equation}
\begin{aligned}
\label{e:GLM}
g(\mu_i) &= \beta_{0} + \sum^{R}_{r=1} Z_{ir}\beta_{r} + \sum^{K}_{k = 1}   \sum^{L}_{l = 1} \zeta_{ikl} \beta_{kl};
\end{aligned}
\end{equation}
where $\beta_{0}$, $\boldsymbol {\beta_{z}} = \{\beta_{r}, r = 1, \cdots, R\}$, $\boldsymbol {\beta_{\zeta}} = \{\beta_{kl}; k = 1, \cdots, K, l = 1, \cdots, L\}$ are the unknown regression coefficients. Note that the number of basis functions $L$ is kept same across $k$ for the purpose of exposition only. We relax this assumption in the next sections assuming $L_k.$ 

In our model, first, we need to estimate $\big\{\zeta_{ikl}, \phi_{k}(\cdot), \psi_{kl}(\cdot); K, L_k \big\}$. Subsequently, using the estimated components, we fit the model \ref{e:GLM} and estimate $\beta_{kl}$'s so as $\gamma.$

\subsection{Selection of basis functions} \label{ss:selectbasis}

The selection of orthonormal basis functions is critical in approximating the model in the form of \ref{e:GLM}. One approach is to select a pre-specified orthonormal basis functions similar to \cite{wang2017classification, fan2015functional, stalling2020}. An alternative approach is to use a data-driven basis; see \cite{staicu2020longitudinal, park2015longitudinal} where the main idea is to estimate the marginal covariance function of the observed functional covariates and select the corresponding orthogonal eigenbasis as a basis. In our approach, we use the latter approach as it has the advantage of exploiting the underlying covariance structures of radiographical images and allows us to account for correlation between the neighboring points.  

The observed functional measurements are most likely to be contaminated with error or white noise. Unlike \cite{wang2017classification}, assume $W_{i}(h, w) = X_{i}(h, w) + \epsilon_{i}(h,w)$, where $W_{i}(\cdot, \cdot)$'s are the noisy realizations of the true functional covariates, $X_{i}(\cdot, \cdot)$'s, and $\epsilon_{i}(\cdot, \cdot)$  is a white noise with mean 0 and covariance, $\sigma_{\epsilon}^{2},$ and is independent over $i,$ $h$, and $w$. Define $\Sigma( w, w') = \int_{\mathcal{H}} E[ X_{i}(h, w) X_{i}(h, w')] g(h) dh$, where $g(\cdot)$ is the sampling density of $h_{u}$'s; $\Sigma( w, w')$ is a proper covariance function (positive semidefinite and symmetric function); see \cite{ramsay2007applied, horvath2012inference}. Define the covariance function of the observed functional covariate realized at a point in $h$-direction by $\Xi(w, w') = \Sigma (w, w') +  \mathds{1} \sigma_{\epsilon}^{2} (w, w').$ The spectral decomposition of $\Xi(w, w')$ results in $\Xi(w, w') = \sum_{k\geq1} \phi_{k}(w) \phi_{k}(w') \lambda_{k};$ where $\lambda_{k}$'s are the non-negative eigenvalues and $\phi_{k}(\cdot)$'s are the orthonormal eigenbasis functions spanned in $L^{2}(\mathcal{W})$ which are used in approximating $\gamma.$  The infnite summation is truncated at a finite level such as $K$ which is determined based on predetermined percentage of variance explained (PVE) value where the main idea is to choose the smallest integer $K$ such that $\sum^{K}_{k=1} \lambda_{k} /\sum^{\infty}_{k=1}\lambda_{k}$ is larger than the preset PVE; such approach is adopted in \cite{goldsmith2011penalized, staicu2020longitudinal, reiss2007functional, di2009multilevel}. This follows that $X_{i}(h, w) \approx \sum^{K}_{k=1} \xi_{ik}(h) \phi_{k}(w)$ and the observed covariate can be written as $W_{i}(h, w) \approx \sum^{K}_{k=1} \xi_{w,ik}(h) \phi_{k}(w),$ where the $k$th observed loadings are $\xi_{w,ik}(h) = \int_{\mathcal{W}} W_{i}(h, w) \phi_{k}(w) dw.$ In particular, $\xi_{w,ik}(h) = \xi_{ik}(h) +  \epsilon_{\xi, h};$  where $\epsilon_{\xi,h}= \int_{\mathcal{W}}  \epsilon_{i}(h, w) \phi_{k}(w) dw$ is  a nugget effect. In a similar spirit to \cite{staicu2020longitudinal}, we model the loadings to exploit the underlying covariance structures in the $h$-direction. Define the covariance function of the $k$th latent smooth process, $\xi_{ik}(\cdot)$ by $\Gamma_{k}(h, h') =  E[\xi_{ik}(h) \xi_{ik}(h')].$ Assume $E[\xi_{ik}(h)] = 0.$  A spectral decomposition of $\Gamma_k(\cdot, \cdot)$ leads to $\Gamma_{k}(h, h') = \sum_{l \geq 1} \psi_{kl}(h) \psi_{kl}(h') \eta_{kl}$, where $\{\psi_{kl}(\cdot), \eta_{kl}\}_{l\geq 1}$'s are the corresponding eigen-components with $\eta_{kl}$'s being the non-negative eigenvalues and  $\psi_{kl}(\cdot)$'s being the orthonormal basis functions spanned in $L^{2}(\mathcal{H}).$ Using the truncated Karhunen-Lo\`eve (KL) expansion, we approximate $\xi_{ik}(\cdot)$ by $\xi_{ik}(h) \approx \sum^{L_{k}}_{l=1} \zeta_{ikl} \psi_{kl}(h);$ where $\zeta_{ikl}$'s are the weights for the $l$th functional principal component (FPC) $\psi_{kl}(\cdot)$ associated with the $k$th direction for the $i$th subject, and is uniquely defined by $\zeta_{ikl} = \int_{\mathcal{H}} \xi_{ik}(h)\psi_{kl}(h) dh$. The $\zeta_{ikl}$'s are the covariates used in modeling \ref{e:GLM}. Such idea of using FPCs with the largest variation are most predictive of responses in an association model is the basis of functional principal component regression (FPCR); \cite{gertheiss2013longitudinal,reiss2007functional}. In a similar spirit to the selection of $K,$ the truncation value $L_k$ is also chosen by prespecified PVE value.

\section{Estimation}  \label{s:estimatebasis}

\subsection{Estimation of components related to functional predictor} Estimation is done following the procedures described in \cite{staicu2020longitudinal, yao2005functional} where the ideas were developed for a sparse design, we borrow the same techniques for a dense design. We briefly review the idea here. We demean the observed functional predictor to ensure $E[W_i(h, w)] = 0.$ Define the demeaned covariates by $\widetilde W(\cdot, \cdot).$ Assume $\widetilde W(\cdot, \cdot)$ is a matrix of $NU \times V$ stacking up all the image data over subjects. A pooled sample covariance estimator is obtained as a cross-product of the elements of $W(\cdot, \cdot)$ divided by the total number of elements in each column such that the $(v, v')$th element is $\widetilde {\Xi}(w_{v}, w_{v'}) = \sum^{N}_{i=1} \widetilde {W}_{i}(h_u, w_v) \widetilde {W}_{i}(h_{u}, w_{v'}) / NU.$ Denote the covariance matrix by $\widetilde {\Xi}(w, w').$  Due to the presence of noise in $\widetilde W(\cdot, \cdot),$ the diagonal terms in $\widetilde \Xi(\cdot, \cdot)$ are inflated by variance and thus requires smoothing. We apply the bivariate smoothing approach \cite{xiao2013fast} to obtain the smoothed covariance function which is denoted by $\widehat \Xi(h, h').$ Applying spectral decomposition on $\widehat {\Xi}(h, h')$ with a pre-set PVE value results in a set of eigen-components, $\{\widehat \lambda_{k}, \widehat \phi_{k}(\cdot)\}^{K}_{k=1}$ where the terms bear the meaning as described in section \ref{ss:selectbasis}. Next the corresponding loadings are obtained through numerical integration as $\widetilde \xi_{w,ik}(h_u) = \int_{\mathcal{W}} \widetilde {W}_{i}(h_u, w) \widehat \phi_{k}(w) dw.$

Let the data obtained from the estimated loadings for a fixed $k$ be $\big\{\widetilde \xi_{w,ik}(h_u); i = 1, \cdots, N; u = 1, \cdots, U\big\}$. Using the idea described in \cite{yao2005functional}, we estimate the covariance function of the loadings and denote it by $\widehat \Gamma_{k}(\cdot, \cdot)$'s. Next the spectral decomposition leads to $\widehat \Gamma_{k}(h, h') \approx \sum^{L_k}_{l = 1} \widehat \psi_{kl}(h) \widehat \psi_{kl}(h') \widehat \eta_{kl}.$ The corresoponding scores $\widehat \zeta_{ikl}$'s are obtained by casting the model for  $\widetilde \xi_{w,ik}(h_u)$'s into a linear mixed model framework assuming that $\widetilde \xi_{w,ik}(\cdot)$'s follow Gaussian distribution; see  \cite{yao2005functional} for details. We replicate this procedure for each $k = 1, \cdots, K$ components.

\subsection{Estimation of response related parameters}
Given the estimated scores $\widehat \zeta_{ikl},$ the approximating model in \ref{e:GLM} can be written as $g(\mu_{i}) = \beta_{0} + \sum^{R}_{r=1} Z_{ir}\beta_{r} + \sum^{K}_{k = 1}   \sum^{L_{k}}_{l = 1} \widehat \zeta_{ikl} \beta_{kl}.$ Define $\beta_{k} = \{\beta_{k1}, \cdots, \beta_{kL}\}$ and let $||\cdot||_{2}$ be the $\ell_{2}$ norms. As in \cite{staicu2020longitudinal} we set PVE generously large enough to capture all non-negligible association between the functional predictor and scalar responses. It is possible that a direction $\phi_{k}(\cdot)$ is only associated with noise and have variance much larger than the signal, such components introduce wiggly pattern of $\gamma$ and are likely to overfit the model. To circumvent this problem, we use a penalized (PEN) technique to estimate parameters. In our application, we also use non-penalized (Non-PEN) version. 


For Non-PEN, we estimate $\boldsymbol{\beta} = \{\beta_{0}, \boldsymbol{\beta^{T}_{z}}, \boldsymbol{\beta^{T}_{\zeta}} \}$ by maximizing the log-likelihood function of the GLM such as
$$ log L(\boldsymbol{\beta}) =  \sum^{N}_{i=1} log  f(y_{i}|X_{i}, Z_{i}; \boldsymbol{\beta})$$  
following the procedures described in \cite{mccullough1989generalized, dobson2018introduction}. 

For PEN, we use a group LASSO penalty to regularize estimation of $\boldsymbol{\beta}$; we impose the penalty on the effects total magnitude $||\beta_{k}||$. Here we maximize the following penalized criterion 
$$ \sum^{N}_{i=1} log  f(y_{i}|X_{i}, Z_{i}; \boldsymbol{\beta}) / N - \kappa \sum^{K}_{k=1} ||\beta_{k}||_{2};$$  
adopting the ideas described in \cite{yuan2006model, yang2015fast}. Here $\kappa \geq 0$ is a tuning parameter controlling sparsity by shrinking all coefficients associated with a group to zero simultaneously; $\kappa$ is selected by K-fold cross-validation (CV) technique.

\subsection{Prediction of response}
Once $\boldsymbol{\beta}$'s are estimated, we estimate the mean of $y_i$ by $\widehat \mu_{i} = g^{-1}( \widehat \beta_{0} + \sum^{R}_{r=1} Z_{ir} \widehat \beta_{r} + \sum^{K}_{k = 1}   \sum^{L_{k}}_{l = 1} \widehat \zeta_{ikl}\widehat \beta_{kl}),$ where $g^{-1}(\cdot)$ is the inverse of the link function $g(\cdot).$ 

For binary classification, we define the predicted responses by $\widehat y_i = 1$ if $\widehat P(y_i = 1) >= c_{opt},$ and $\widehat y_i = 0$ otherwise. Here, $c_{opt}$ is the optimum cut-off value based on Youden index and optimized by maximizing the sum of sensitivity and specificity; see \cite{zweig1993receiver, shapiro1999interpretation, greiner2000principles}.

\section{Inference}  \label{s:inference}
Though our idea is illustrated by using 2D X-ray images, the proposed methodology can be leveraged in other radiography images such as magnetic resonance imaging (MRI), positron emission tomography (PET) images, ultrasounds, and mammography. While there are various kinds of medical images, it is a question of interest to investigate which images are sufficiently discriminative enough to differentiate between cases and non-cases. For instance, if there is no difference between the chest X-rays of pneumonia and healthy patients, then the odds ratio for a given covariate $Z_i$ will be equal to 1. Such a hypothesis of interest can be written formally as
%
%
\begin{equation}
\label{Main Hypo1}
\begin{aligned}
H_{0} &:  \gamma(h, w) = 0, \hspace{0.2cm} \text{for all} \hspace{0.1cm} \text{h} \hspace{0.1cm} \text{and} \hspace{0.1cm} \text{w},\\
H_{A} &:  \gamma(h, w) \neq 0, \hspace{0.2cm} \text{for some} \hspace{0.1cm} \text{h or w}.
\end{aligned}
\end{equation} 
Equivalently, with an abuse of notation, we write $||\gamma||^{2} = 0,$ where  $||\gamma||^{2} = \int_{\mathcal{H}} \int_{\mathcal{W}} \gamma^{2}(h, w) dw dh.$ Using the orthonormal property of the basis functions $\{\phi_{k}(\cdot), \psi_{kl}(\cdot); k = 1, \cdot, K, l = 1, \cdots, L_{k}\},$ this can further be written as 
\begin{equation*} 
\begin{aligned}
0 =  ||\gamma||^{2} &= \int_{\mathcal{H}} \int_{\mathcal{W}}  \left[ \left\{ \sum_{k \geq 1} \sum_{l \geq 1} \phi_{k}(w)  \psi_{kl}(h) \beta_{kl} \right\} \left\{ \sum_{k' \geq 1} \sum_{l' \geq 1} \phi_{k'}(w)  \psi_{k'l'}(h) \beta_{k'l'} \right\}  \right] dh dw \\
&=  \sum_{k \geq 1} \sum_{l \geq 1} \beta^{2}_{kl}. 
\end{aligned} 
\end{equation*}  
This follows that the hypothesis \ref{Main Hypo1} can be reformulated as 
\begin{equation}
\label{Main Hypo2}
\begin{aligned}
H_{0} &:  \beta_{kl} = 0, \hspace{0.2cm} \text{for all} \hspace{0.1cm} \text{k and l},\\
H_{A} &:  \beta_{kl} \neq 0, \hspace{0.2cm} \text{for at-least one k or l}.
\end{aligned}
\end{equation}   
The parsimonious modeling framework \ref{e:GLM} allows to make inference about the significance of the association between binary responses and functional measurements through testing the nullity of $\beta_{kl}$'s for all $k$'s and $l$'s using the conventional hypothesis testing procedures such as likelihood ratio test (LRT), F-1 test, Wald test, or score test. While all these tests are asymptotically equivalent, we use LRT due to its amenable theoretical properties in testing both constrained and unconstrained parameter space; see \cite{johnston1963econometric, molenberghs2007likelihood}. In particular, the test is based on the differences between the likelihoods computed from the maximum likelihood estimates (MLEs) under $H_{0}$ and $H_{A}.$ Define the test statistic by
\begin{equation}
\label{test statistic}
\mathcal{T}  = - 2 \Big\{log L(\boldsymbol{\widetilde \beta}) - log L(\boldsymbol{\widehat \beta}) \Big\},
\end{equation}
where $L(\boldsymbol{\widetilde \beta})$ refers to the likelihood value with respect to the MLEs under $H_{0}$ and $L(\boldsymbol{\widehat \beta})$ the MLEs under $H_{A}$; it can be shown that $\mathcal{T}$ follows asympotically $\chi^{2}$ distribution with $\nu$ degrees of freedom where $\nu$ equals the difference in the number of free parameters to be estimated under $H_{0}$ and $H_{A}$ such that $\nu = \sum^{K}_{k=1}L_{k}.$ LRT has been studied extensively and implemented in various applications for decades; we refer to \cite{peugh2010practical, wood2011fast, mccullagh2018generalized} and references there in for details. Note that such hypothesis testing procedure makes more sense for the Non-PEN model. If the PEN model is used, $\tau$ will ideally shrink all coefficents, $\beta_{k}$'s, to zero when there is no association between responses and covariates given that the optimum value for $\tau$ is chosen appropriately.  



\section{Pneumonia image classification}\label{s:data analysis}
We apply the methodology on the pediatric pneumonia dataset described in section \ref{s:data}. The main objectives are to classify the images by \textit{(S1)} regular and pneumonia and \textit{(S2)} viral pneumonia and bacterial pneumonia cases, and evaluate the statistical significance of the association between X-ray images and binary classes. In particular, we use the model
\begin{equation}
\begin{aligned}
\label{e:FLM}
g(\mu_i) &= \beta_{0} + \int_\mathcal{H} \int_\mathcal{W} X_{i}(h, w) \gamma(h, w) dw dh;
\end{aligned}
\end{equation} 
where $g(\cdot)$ is a logit link assuming $y_i$ is a binary response such that in \textit{(S1)}, $y_i = 1$ for pneumonia case and 0 otherwise. Similarly in \textit{(S2)}, let be $y_i = 1$ for bacterial pneumonia, and 0 for viral pneumonia. While we use both healthy and pneumonia data in the former classification problem, the latter exploits only the pneumonia data.

\subsection{Computational details} We implement the method in R version 3.6.3 \cite{team2013r}. We read and edit the X-ray images using the package {\tt EBImage} \cite{pau2010ebimage}. We use the {\tt fpca.face} and {\tt fpca.sc} functions from the {\tt refund} \cite{goldsmith5refund} to estimate the covariates related components. We set the PVE value equal to 0.99 in our application. The Non-PEN is solved using the {\tt glm} function from the package {\tt stats}. The PEN model is fitted by {\tt gglasso} function of the {\tt gglasso} package \cite{yang2015fast}. The LRT is calculated using the {\tt lrtest} function from the {\tt lmtest} package \cite{lmtest}. The computation time to train and test the Non-PEN model in a sample of 2,600 X-ray images is approximately 40 seconds, and it takes approximately 240 seconds to train-and-validate the sample of 5,216 images in a machine with an Intel Core-i7 processor having 16GB RAM. 

\subsection{Permormance metrics} We split the data into a training set (in-sample) on which the model is built and a test set (out-of-sample) on which the performance is evaluated; we report the results for both in-sample and out-of-sample dataset. Two sampling mechanisms are considered. \textit{(M1)} All 5,216 images (1,341 standard and 3,875 pneumonia) are used where 350 pictures from each type are selected randomly and kept aside to constitute the test dataset; the remaining 4,466 X-rays are used to train the model. \textit{(M2)} A random sample of 1,300 from each type (i.e., healthy and pneumonia) so that a total of 2,600 images form the whole dataset. As in \textit{(M1)}, a total of 700 copies (i.e., 350 healthy and 350 pneumonia) form the test set, and the remaining 1,900 images build the training cohort. There are 1,345 viral and 2,345 bacterial pneumonia X-rays. Notice \textit{(M2)} ensures balanced cases and uses much smaller sample to train the model than that of \textit{(M1)}; the purpose is to assess the model's performance with a smaller sample. We consider two ways of forming the test set - \textit{(a)} replicating the test-train split 200 times and evaluating the performance within each split \textit{(b)} forming the test-train split for once. In both cases, test images are selected randomly without replacement. Approach \textit{(a)} will potentially induce more randomness into the train-test split and is less susceptible to sampling bias. Let $\mathcal{N}_p$ and $\mathcal{N}_h$ be the set of patients referring to the pneumonia (i.e., $y_i = 1$) and healthy cases (i.e., $y_i = 0$) respectively such that $|\mathcal{N}_p| + |\mathcal{N}_h| = N,$ where $|\cdot|$ is the cardinality of a set. Define 
\begin{itemize}
 \item True positive (TP) = $\sum^{N}_{i=1} \widehat y_i \mathbbm{1} \big(i \in \mathcal{N}_p \big)$
 \item False negative (FN) = $\sum^{N}_{i=1} (1 - \widehat y_i) \mathbbm{1} \big(i \in \mathcal{N}_p \big)$
\item True negative (TN) = $\sum^{N}_{i=1} (1 - \widehat y_i) \mathbbm{1} \big(i \in \mathcal{N}_h \big)$
\item False positive (FP) = $\sum^{N}_{i=1} \widehat y_i \mathbbm{1} \big(i \in \mathcal{N}_h \big)$
\end{itemize}
Here $\mathbbm{1}(\cdot)$ is an indicator function taking values 0 and 1. We consider the following performance metrics
\begin{itemize}
 \item True positive rate (TPR) = TP / (TP + FN)
\item True negative rate (TNR) = TN / (TN + FP)
 \item Positive predictive value (PPV) = TP / (TP + FP)
 \item Accuracy = TP + TN / (TP + TN + FP + FN)
\item Mathhews correlation coefficient (MCC) = $(\text{TP} \cdot \text{TN} - \text{FP} \cdot \text{FN}) / \\ \sqrt{ \big\{ (\text{TP} + \text{FP}) \cdot (\text{TP} + \text{FN}) \cdot (\text{TN} + \text{FP}) \cdot (\text{TN} + \text{FN}) \big\} }$
\item F$_1$ = $2 \cdot \text{TPR}  \cdot \text{PPV} / (\text{PPV} + \text{TPR})$
\item Area under the curve  (AUC) of a receiver operator characteristic (ROC) curve; ROC is generated by plotting the TPR versus false positive rate (FPR) at various thresholds of a binary classifier system.
\end{itemize} 

\subsection{Classification performance}
Table \ref{class_metrics1} provides the numerical performance of the model for both in-sample and out-of-sample image data for \textit{(S1)}. The classification results are close in the validation set for both \textit{(a)} and \textit{(b)}; due to the smaller training sample in \textit{(b)}, unsurprisingly, in-sample performances are marginally better. The high ACC values indicate that the model identifies pneumonia and healthy patients using X-rays correctly. The high F$_1$ values suggest a good balance between precision (i.e., TPR) and recall (i.e., PPV) referring to the number of instances that the model correctly classifies and instances that the model misses, respectively. Besides, the high AUC values ($ \geq 0.98$) indicate an excellent balance between TPR and the false positive rate (FPR). Apparently, we observe better classification performances in the Non-PEN method for \textit{(M1)} than that of \textit{(M2)}. This could be due to the fact that \textit{(M1)} is comprised of all images implying more variation in the data; therefore, an FPC analysis with a high PVE (i.e., 0.99) is likely to provide some directions associated with noise that could potentially overfit the model. Therefore, by shrinking such directions to zero, we improve the model performance. The IQR values provide evidence of small variation in the reported summary values across 200 validation sets for \textit{(a)}. The other important metrics such as TNR, PPV, and $F_1$ are reported in the Supplementary Material; see Section  \label{s:Image} in \ref{s:sup}. 

Table \ref{class_metrics2} provides the numerical performance of the model for \textit{(S2)}. Separating viral and bacterial cases using a scalar-on-image regression is a much more difficult problem than that of \textit{(S1)}. Overall, the model performs satisfactorily in differentiating between viral and bacterial pneumonia patients. As above, we observe better out-of-sample performances in PEN approach. 

We are interested in evaluating two scientific questions. (1) Do the radiographic image data differ between healthy and pneumonia patients? (2) Do the image data differ between viral and pneumonia patients? We use \ref{test statistic} to test the hypotheses. As the calculated p-value is $<$ \text{2.2e-16} for testing (1), we may conjecture that the observed patterns in the X-rays are distinctive between healthy individuals and pneumonia patients. Similarly, since the p-value for testing (2) is $<$ \text{2.2e-16}, we may conclude that there is a statistically significant difference in the image data between viral and pneumonia patients. Further results related to testing the equality of mean scores between healthy and pneumonia patients are provided in Section  \label{s:Image} of the Supplementary Material \ref{s:sup}.
\begin{table}[ht]
\tiny
\caption{Comparison of chest X-rays between pneumonia and normal \textit{(S1)}. Classification metrics are presented for in-sample and out-of-sample data with respect to training-testing schemes \textit{(a)} and \textit{(b)} based on \textit{(M1)} and \textit{(M2)}. Median and corresponding IQR (in parenthesis) across 200 iterations are reported for scheme \textit{(a)}.}
\label{class_metrics1}
\noindent\makebox[\textwidth]{ 
\begin{tabular}{ccccccc}
\multicolumn{1}{l}{} & \multicolumn{3}{c}{Non-PEN} & \multicolumn{3}{c}{PEN} \\ \cline{2-7} 
Settings & ACC & TPR & AUC & ACC & TPR & AUC \\ \cline{2-7} 
M1 + (a) + In & 0.950 {[}0.006{]} & 0.952 {[}0.010{]} & 0.986 {[}0.001{]} & 0.963 {[}0.006{]} & 0.962 {[}0.010{]} & 0.994 {[}0.001{]} \\ \cline{2-7} 
M1 + (a) + Out & 0.931 {[}0.010{]} & 0.931 {[}0.023{]} & 0.979 {[}0.005{]} & 0.949 {[}0.010{]} & 0.957 {[}0.017{]} & 0.988 {[}0.004{]} \\ \cline{2-7} 
M1 + (b) + In & 0.957 & 0.962 & 0.987 & 0.972 & 0.974 & 0.995 \\ \cline{2-7} 
M1 + (b) + Out & 0.926 & 0.940 & 0.976 & 0.943 & 0.972 & 0.983 \\ \cline{2-7} 
M2 + (a) + In & 0.957 {[}0.004{]} & 0.952 {[}0.014{]} & 0.992 {[}0.001{]} & 0.956 {[}0.004{]} & 0.954 {[}0.013{]} & 0.991 {[}0.001{]} \\ \cline{2-7} 
M2 + (a) + Out & 0.940 {[}0.010{]} & 0.934 {[}0.022{]} & 0.986 {[}0.004{]} & 0.941 {[}0.010{]} & 0.940 {[}0.020{]} & 0.987 {[}0.004{]} \\ \cline{2-7} 
M2 + (b) + In & 0.956 & 0.945 & 0.991 & 0.957 & 0.955 & 0.991 \\ \cline{2-7} 
M1 + (b) + Out & 0.943 & 0.943 & 0.988 & 0.940 & 0.946 & 0.987 \\ \cline{2-7} 
\end{tabular}
}
\end{table}

\begin{table}[ht]
\tiny
\caption{Comparison of chest X-rays between viral and bacterial pneumonia \textit{(S2)}. Classification metrics are presented for in-sample and out-of-sample data with respect to training-testing schemes \textit{(a)} and \textit{(b)} based on \textit{(M1)} and \textit{(M2)}. Median and corresponding IQR (in parenthesis) across 200 iterations are reported for scheme \textit{(a)}.}
\label{class_metrics2}
\noindent\makebox[\textwidth]{ 
\begin{tabular}{ccccccc}
 & \multicolumn{3}{c}{Non-PEN} & \multicolumn{3}{c}{PEN} \\ \cline{2-7} 
Settings & ACC & TPR & AUC & ACC & TPR & AUC \\ \cline{2-7} 
M1 + (a) + In & 0.728 {[}0.021{]} & 0.717 {[}0.056{]} & 0.807 {[}0.006{]} & 0.731 {[}0.014{]} & 0.739 {[}0.036{]} & 0.795 {[}0.007{]} \\ \cline{2-7} 
M1 + (a) + Out & 0.690 {[}0.023{]} & 0.674 {[}0.023{]} & 0.759 {[}0.022{]} & 0.701 {[}0.021{]} & 0.729 {[}0.054{]} & 0.768 {[}0.024{]} \\ \cline{2-7} 
M1 + (b) + In & 0.736 & 0.731 & 0.810 & 0.723 & 0.707 & 0.800 \\ \cline{2-7} 
M1 + (b) + Out & 0.689 & 0.700 & 0.753 & 0.689 & 0.691 & 0.756 \\ \cline{2-7} 
M2 + (a) + In & 0.754 {[}0.009{]} & 0.743 {[}0.047{]} & 0.827 {[}0.007{]} & 0.721 {[}0.009{]} & 0.748 {[}0.059{]} & 0.791 {[}0.010{]} \\ \cline{2-7} 
M2 + (a) + Out & 0.686 {[}0.022{]} & 0.677 {[}0.059{]} & 0.746 {[}0.021{]} & 0.693 {[}0.020{]} & 0.721 {[}0.078{]} & 0.760 {[}0.020{]} \\ \cline{2-7} 
M2 + (b) + In & 0.761 & 0.738 & 0.828 & 0.728 & 0.777 & 0.792 \\ \cline{2-7} 
M2 + (b) + In & 0.693 & 0.675 & 0.740 & 0.696 & 0.726 & 0.763 \\ \cline{2-7} 
\end{tabular}
}
\end{table}


\section{Numerical experiment}\label{s:simulation}
\subsection{Classification performance on simulated data}
\begin{figure}[ht]
	\centering
	\caption{Example of simulated chest X-ray images for a healthy subject (top) and a patient with pneumonia (bottom) with $\sigma \in \{0.01, 0.10, 0.20, 0.50, 1.00\}$ from left-to-right in an increasing order.}
\label{f:xray_simul} 
     \includegraphics[width=0.18\textwidth]{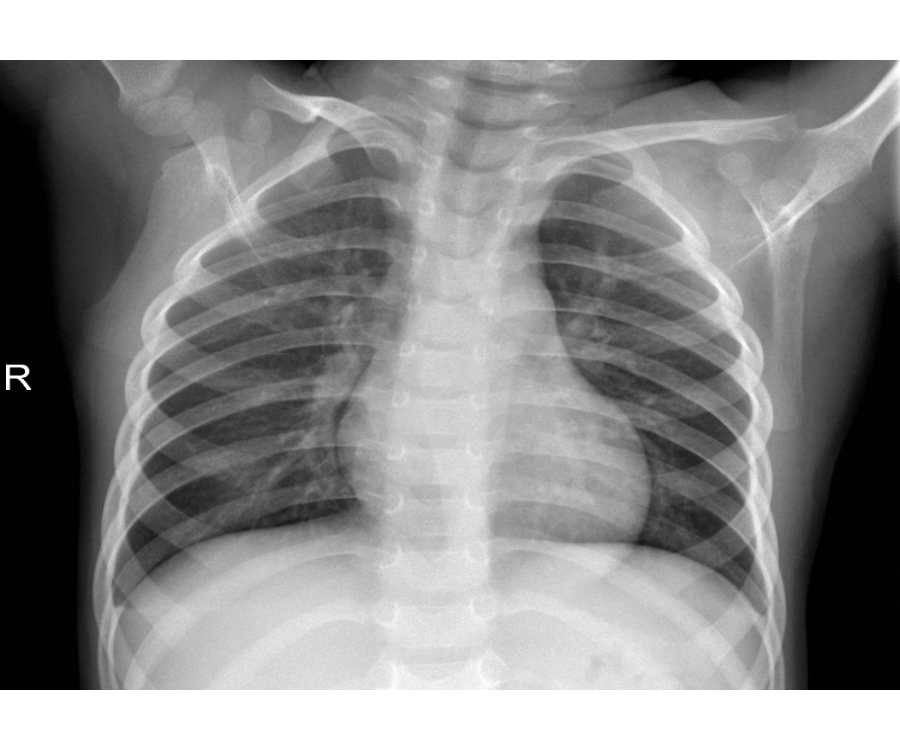}
      \includegraphics[width=0.18\textwidth]{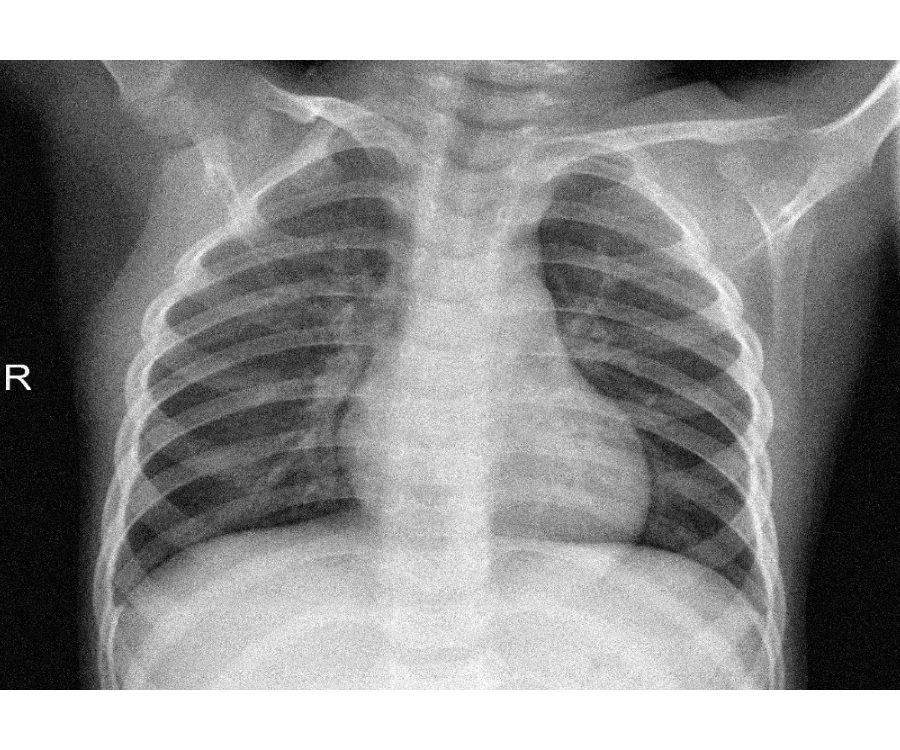}
	\includegraphics[width=0.18\textwidth]{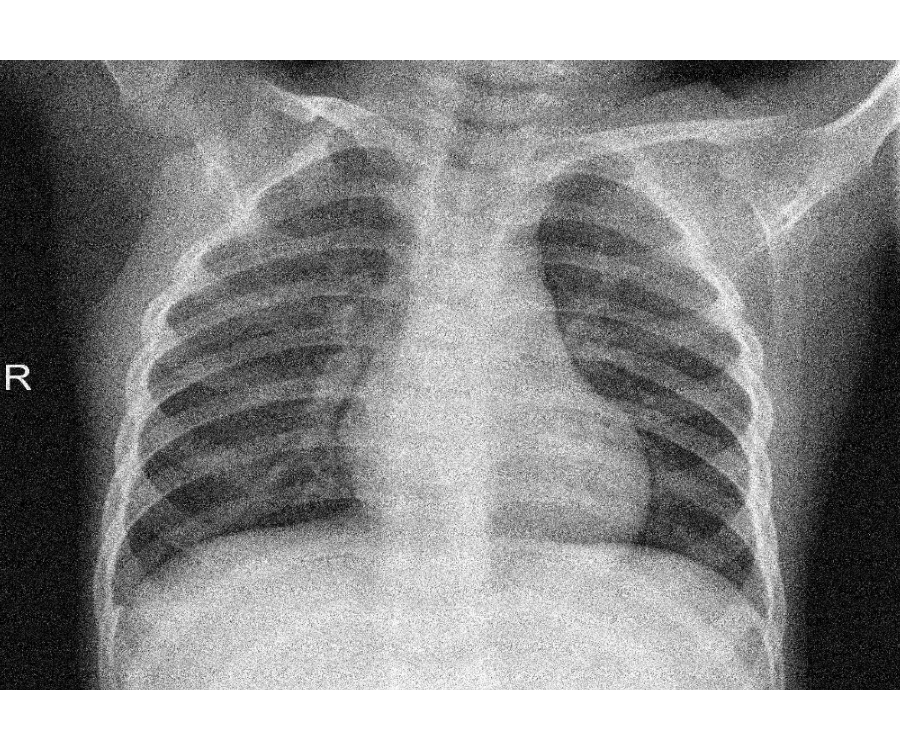}
	\includegraphics[width=0.18\textwidth]{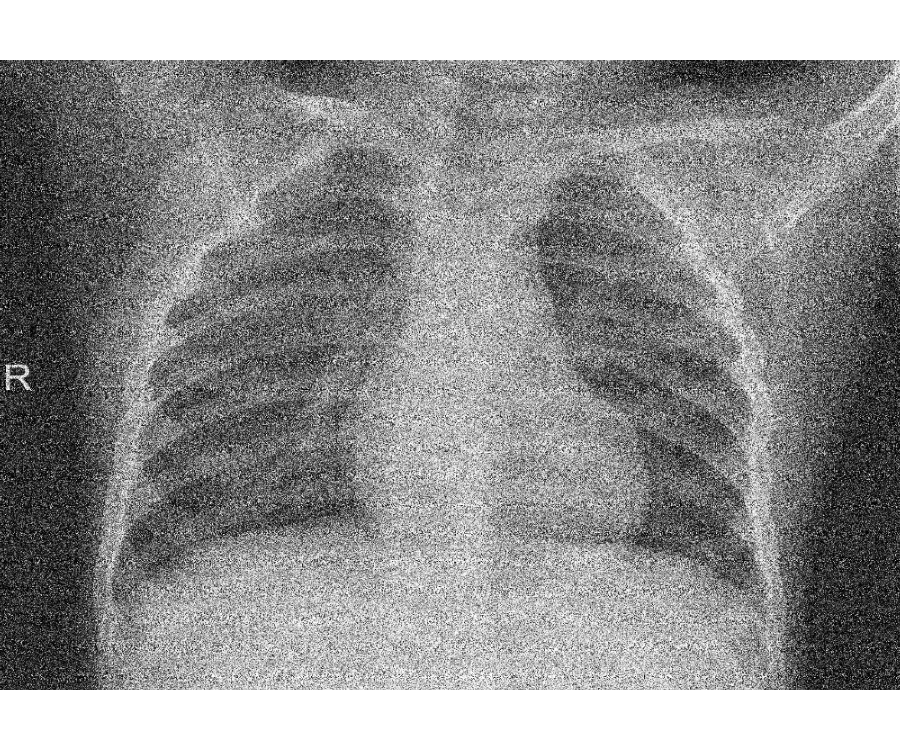}
 	\includegraphics[width=0.18\textwidth]{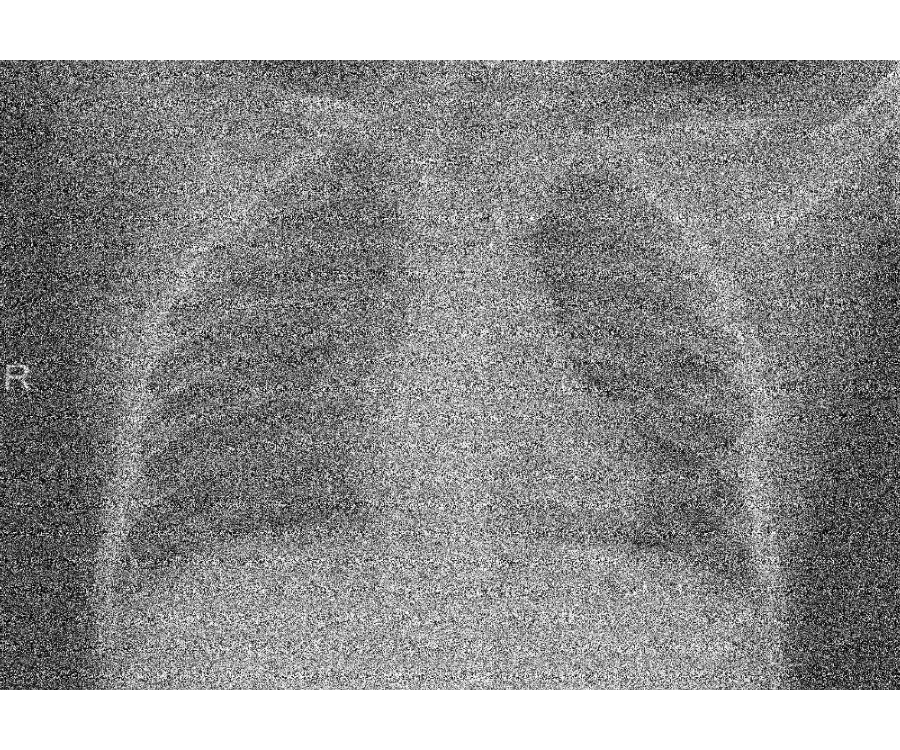}
      \includegraphics[width=0.18\textwidth]{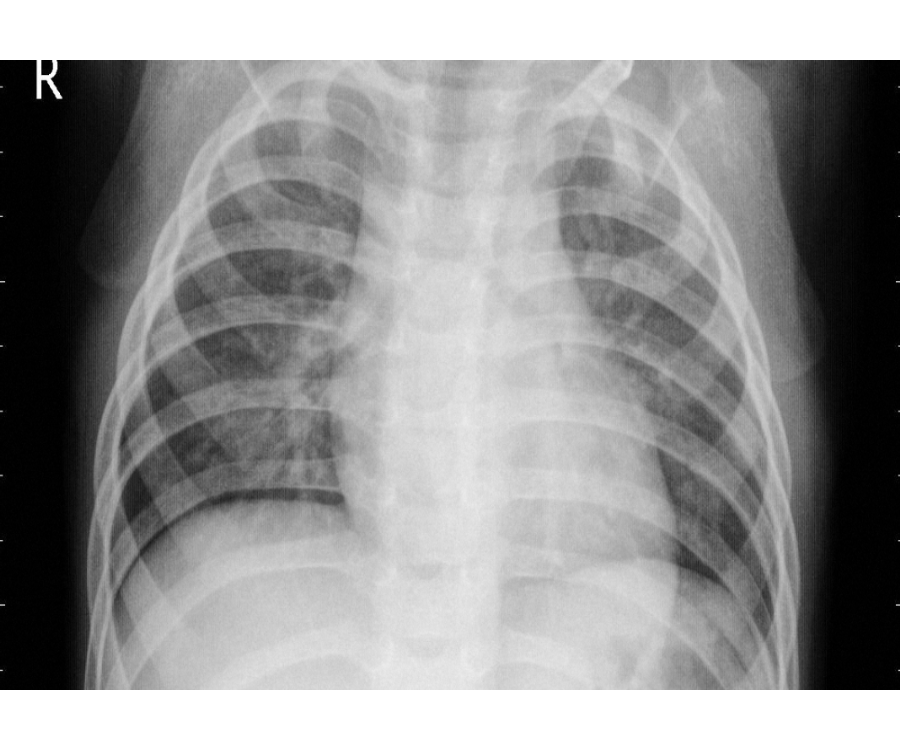}
	\includegraphics[width=0.18\textwidth]{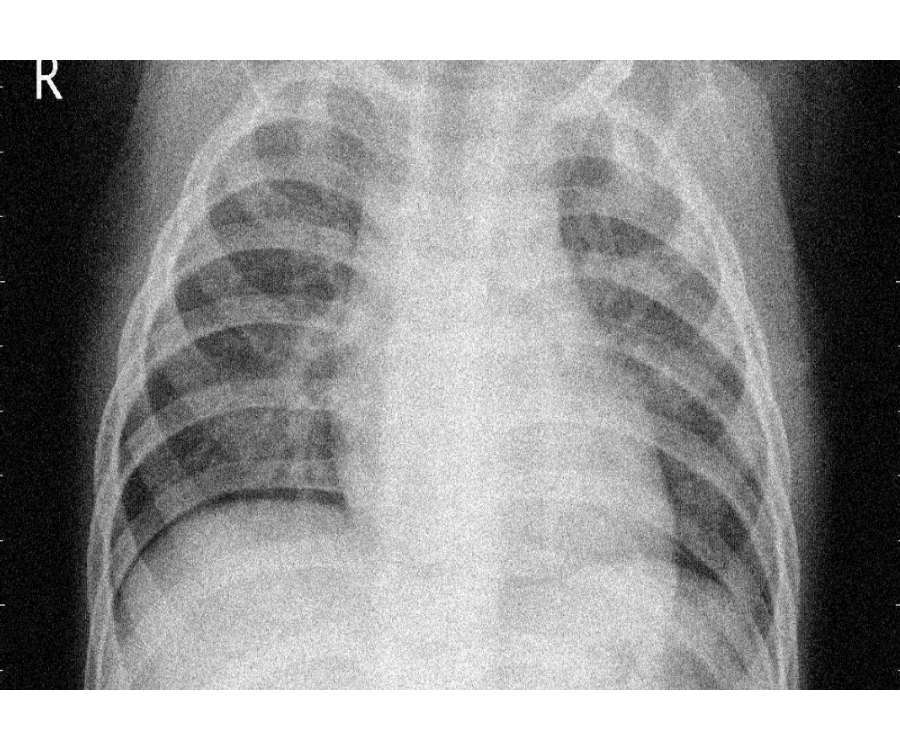}
	\includegraphics[width=0.18\textwidth]{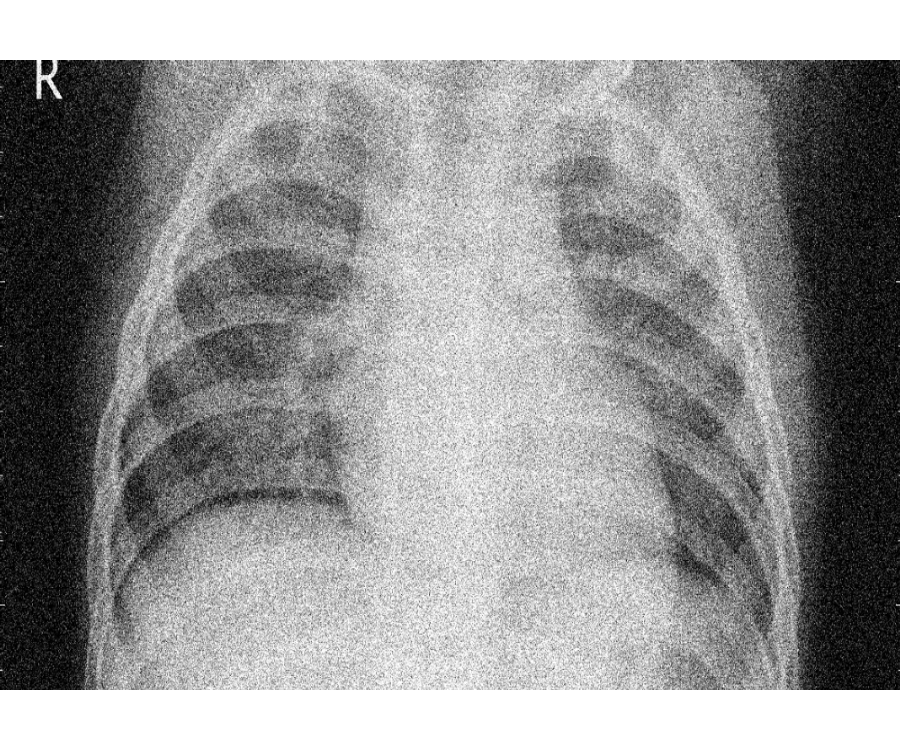}
	\includegraphics[width=0.18\textwidth]{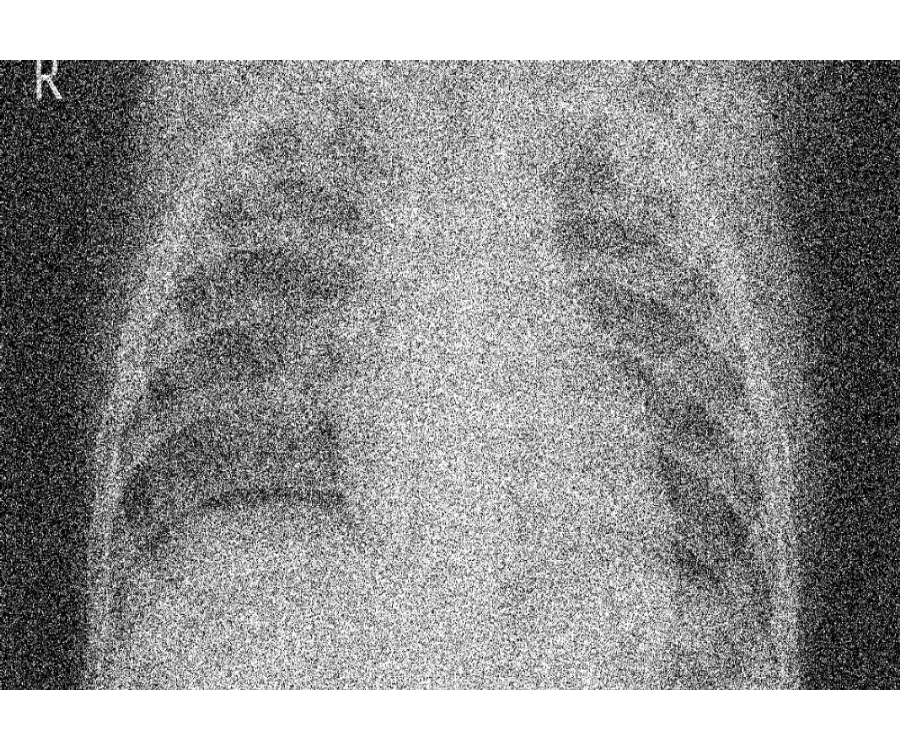}
	\includegraphics[width=0.18\textwidth]{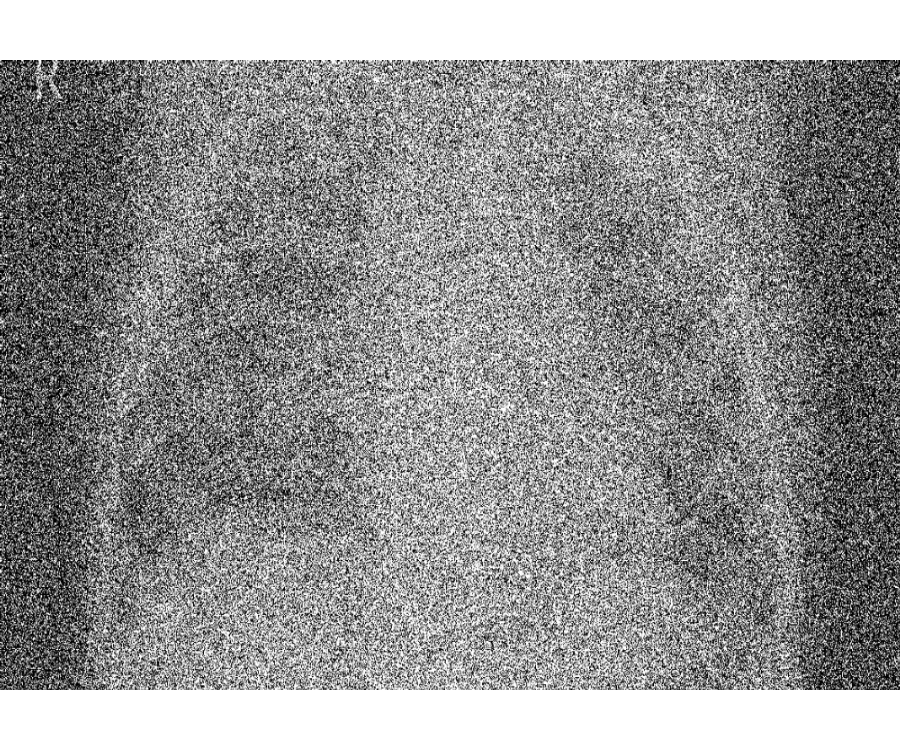}
\end{figure}

In this section, we consider a simulation study generating X-ray images for healthy and pneumonia patients. The objective is to assess the model performance when the images are alterted by noises. Let the observed data be  $\big[y_{i^{*}},  \{X_{i^{*}}(h_{u},w_{v}); u = 1, \ldots, U, v = 1, \ldots, V\}; i^{*} = 1, \ldots, N^{*}\big];$ where $y_{i^{*}}$ is the binary response associated with the $i^{*}$th subject for who the simulated image, $X_{i^{*}}(\cdot,\cdot).,$ is rendered.  Let $N^{*} = 600.$ We generate gray-scaled $120 \times 80$ pixel images for each patient such that $U = 120$ and $V = 80.$ Denote by  $X_{i^{*}}(h_{u}, w_{v})  = X_{i}(h_{u}, w_{v}) + \varepsilon_{uv}$ the model that generates noisy realizations of X-ray images; where $\varepsilon_{uv}$ is a white noise which follows IID $\pN (0, \sigma^{2}).$ Here $ X_i(\cdot, \cdot)$'s are the original gray-scaled images selected randomly from 5,216 original images; i.e. $i \in \mathcal{N}_{h}$ or $i \in \mathcal{N}_{p}$. Note that the standard deviation $\sigma$ controls the departure in pixel from the true X-ray images; as $\sigma$ departs from zero, the image quality drops gradually and gets perturbed on both $u$ and $v$ directions. In our simulation study, we consider $\sigma \in \{1\text{e-8}, 0.01, 0.10, 0.20, 0.50, 1.00\};$ see Figure \ref{f:xray_simul}. Note that as the magnitude of noise increases, it gets difficult to interpret and becomes indistinguishable between healthy and pneumonia cases. For each setting, we run simulations for 500 times, and within each run, we simulate the X-ray images for 300 healthy and 300 pneumonia patients. We label the binary responses such that $y_{i^{*}} = 0$ if a patient belongs to the healthy cohort and 1, otherwise. To assess the classification performance of the proposed method, we divide each simulated dataset into training (75\%) and test set (25\%) and calculate the classification metrics for both in-sample and out-of-sample data. We set PVE equal to 0.90 for this experiment.

Table \ref{simul_table} displays the binary classification results for the simulated data with different magnitude of noises. The results are close for $\sigma = 0.01$ and $\sigma = 0.10$, which provides evidence of the efficacy of the proposed methodology in the presence of noise. Overall, both PEN and Non-PEN perform competitively. While the model still classifies patients satisfactorily for $\sigma \geq 0.50,$ the performance drops to an extent. This phenomenon is not too unexpected as it becomes hard for the model to extract the signal out of the data from such noisy images. In reality, such low-quality images are not even recommended for clinical use anyway. Additional results are provided in Section \label{s:Simul} of the Supplementary Material \ref{s:sup}.

Figure \ref{f:xray_auc} illustrates the area under the curve (AUC) of a receiver operator characteristic for different noise levels for the Non-PEN (left) and PEN (right) model; the differences are very marginal and visually indistinguishable. The model performance with respect to AUC remains somewhat similar for $\sigma = 0.01$ and $\sigma = 0.10;$ As expected, with $\sigma > 0.20$, the performance of the classification model decays gradually.

\begin{table}[ht]
\tiny
\caption{Comparison of chest X-rays between simulated pneumonia and normal \textit{(S1)} for balanced cases \textit{(b)}. Classification metrics (ACC, TPR, TNR) for in-sample and out-of-sample data at $\sigma \in \{0.01, 0.10, 0.20, 0.50, 1.00\}$ are summarized for Non-PEN and PEN estimation approach. Median values and the corresponding IQR (in parenthesis) are reported. Results are based on 500 simulations.}
\label{simul_table}
\noindent\makebox[\textwidth]{ 
\begin{tabular}{cccccccc}
\multicolumn{1}{l}{} & \multicolumn{1}{l}{} & \multicolumn{3}{c}{Non-PEN} & \multicolumn{3}{c}{PEN} \\ \cline{3-8} 
$\sigma$ & Validation & ACC & TPR & TNR & ACC & TPR & TNR \\ \cline{3-8} 
$0.01$ & In & 0.933 {[}0.011{]} & 0.920 {[}0.036{]} & 0.947 {[}0.027{]} & 0.931 {[}0.011{]} & 0.916 {[}0.036{]} & 0.947 {[}0.031{]} \\ \cline{3-8} 
 & Out & 0.900 {[}0.033{]} & 0.893 {[}0.053{]} & 0.920 {[}0.053{]} & 0.907 {[}0.033{]} & 0.893 {[}0.067{]} & 0.920 {[}0.053{]} \\ \cline{3-8} 
$0.10$ & In & 0.931 {[}0.011{]} & 0.920 {[}0.031{]} & 0.942 {[}0.027{]} & 0.929 {[}0.011{]} & 0.916 {[}0.036{]} & 0.942 {[}0.031{]} \\ \cline{3-8} 
 & Out & 0.900 {[}0.027{]} & 0.893 {[}0.067{]} & 0.920 {[}0.053{]} & 0.900 {[}0.027{]} & 0.887 {[}0.067{]} & 0.920 {[}0.053{]} \\ \cline{3-8} 
$0.20$ & In & 0.899 {[}0.013{]} & 0.890 {[}0.031{]} & 0.911 {[}0.040{]} & 0.898 {[}0.013{]} & 0.889 {[}0.027{]} & 0.907 {[}0.036{]} \\ \cline{3-8} 
 & Out & 0.873 {[}0.033{]} & 0.853 {[}0.067{]} & 0.893 {[}0.067{]} & 0.873 {[}0.033{]} & 0.867 {[}0.067{]} & 0.893 {[}0.067{]} \\ \cline{3-8} 
$0.50$ & In & 0.807 {[}0.018{]} & 0.804 {[}0.062{]} & 0.813 {[}0.063{]} & 0.809 {[}0.020{]} & 0.809 {[}0.058{]} & 0.809 {[}0.062{]} \\ \cline{3-8} 
 & Out & 0.787 {[}0.040{]} & 0.787 {[}0.080{]} & 0.787 {[}0.080{]} & 0.787 {[}0.040{]} & 0.787 {[}0.080{]} & 0.787 {[}0.080{]} \\ \cline{3-8} 
$1.00$ & In & 0.780 {[}0.018{]} & 0.773 {[}0.072{]} & 0.796 {[}0.071{]} & 0.780 {[}0.018{]} & 0.773 {[}0.076{]} & 0.796 {[}0.080{]} \\ \cline{3-8} 
 & Out & 0.753 {[}0.042{]} & 0.747 {[}0.107{]} & 0.773 {[}0.093{]} & 0.753 {[}0.047{]} & \multicolumn{1}{l}{0.747 {[}0.097{]}} & \multicolumn{1}{l}{0.773 {[}0.093{]}} \\ \cline{3-8} 
\end{tabular}
}
\end{table}

\begin{figure}[h]
	\centering
	\caption{Boxplots of AUC values estimated by Non-PEN (left) and PEN (right) model for different noise levels $\sigma \in \{0.01, 0.10, 0.20, 0.50, 1.00\}.$ Median values, 25th, and 75th percentile values are displayed along with upper and lower whiskers. The results are based on 500 simulations.}
\noindent\makebox[\textwidth]{ 
	\includegraphics[width=0.45\textwidth]{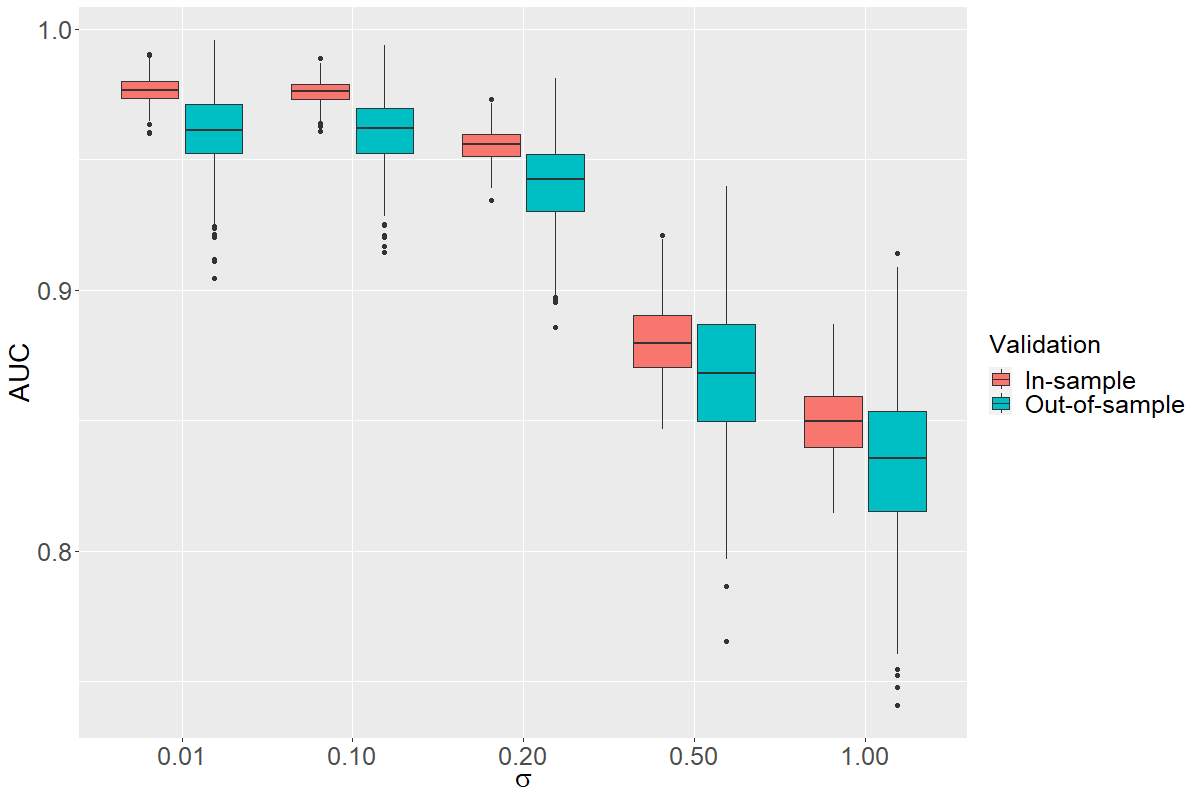} 
       \includegraphics[width=0.45\textwidth]{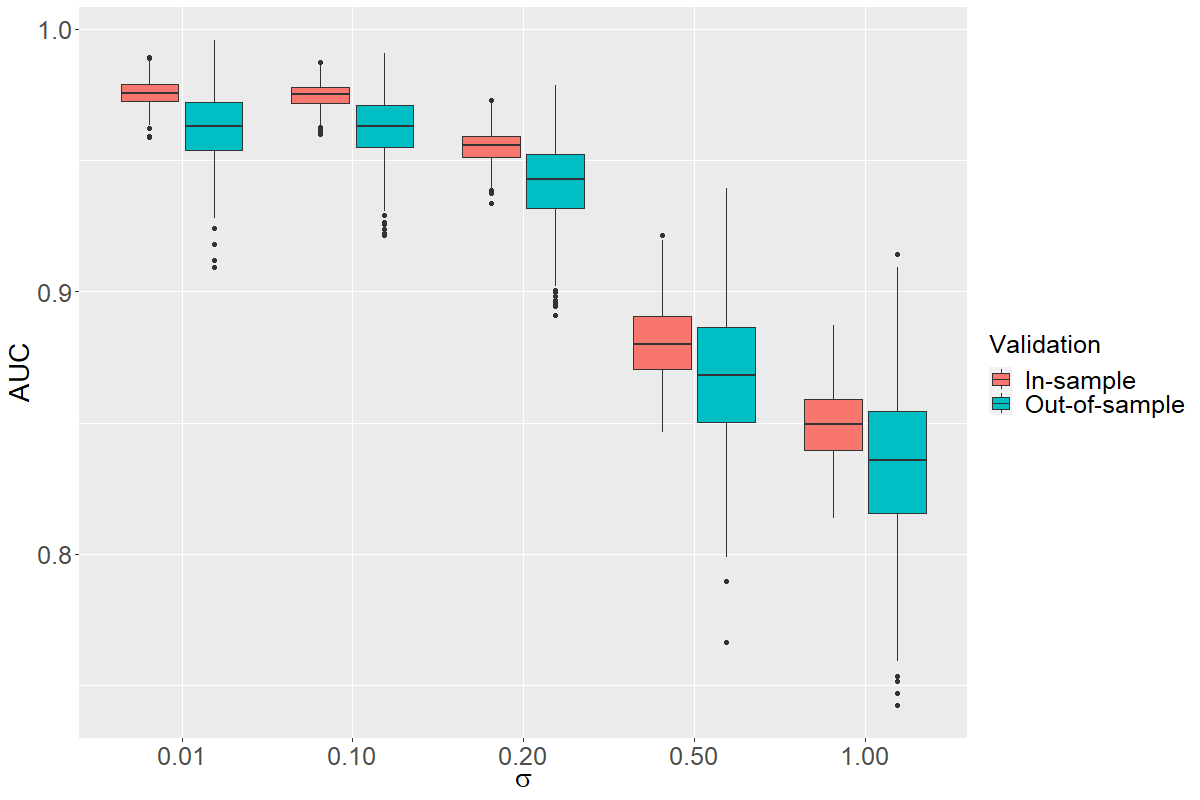}
}
	\label{f:xray_auc} 
\end{figure}

\subsection{Numerial evaluation of inferential property}


As before, we simulate the images by  $X_{i^{*}}(h_{u}, w_{v})  = X_{i}(h_{u}, w_{v}) +  \varepsilon_{uv},$ where $i \in \mathcal{N}_{h}$ or $i \in \mathcal{N}_{p}$, and $\varepsilon_{uv}$ follows Gaussian distribution with mean 0 and standard deviation $\sigma \in \{1e\text{e-8}, 0.01, 0.10\},$ and is an IID. Let $N^{*} = 600.$ All $X_{i^{*}}(\cdot, \cdot)$'s are selected randomly from the pool of original 5,216 images. Let  $\rho$ control the number of samples taken from the pneumonia cohort. We consider two scenarios. \textit{(N1)} Simulate 600 $X_{i}(\cdot, \cdot)$'s from the original 1,341 files associated with the normal X-rays by sampling without replacement. We assign the first 300 labels as noncases (i.e., $y_{i^{*}} = 0$) and the rest 300 as cases (i.e.,  $y_{i^{*}} = 1$) ignoring their true class assignments as $i \in \mathcal{N}_{h}.$ \textit{(N2)} Simulate $X_{i}(\cdot, \cdot)$'s in such a way so that data are comprised of both normal and penumonia patients. The first 300 X-rays are generated from the healthy cohort (i.e., $i \in \mathcal{N}_{h}$) and we label the corresponding responses as $y_{i^{*}} = 0.$ For the second half, $(1 - \rho)$\% of the remaining 300 images are selected from the healthy cohort (i.e., $i \in \mathcal{N}_{h}$) and $\rho$\% are from the pneumonia cohort (i.e., $i \in \mathcal{N}_{p}$). For these 300 cases, we label the corresponding responses as  $y_{i^{*}} = 1.$ Note that \textit{(N1)} and \textit{(N2)} evaluate the rejection rates of $\mathcal{T}$ under $H_{0}$ and  $H_{A}$ as in \ref{Main Hypo1}, respectively. 

The top section of table \ref{test_table} reports the rejection probabilities and corresponding standard errors for testing the null effect under the case when indeed $H_{0}$ is true. The results are based on 8,000 simulations. The empirical type-I error rates are around the nominal levels for $\sigma \in \{1\text{e-8}, 0.01, 0.10\}.$ However, we notice slightly lower rejection rates for $\sigma = 0.01$. The bottom section illustrates the power properties for testing the null effect when data are generated under $H_{A}$. The results associated with the power are based on 500 simulations. As $\rho$ departs from 0, more pneumonia patients are added in the sample. As expected, the empirical rejection rates for the model get closer to 1 for higher $\rho$ at different $\sigma.$
\begin{table}[ht]
\tiny
\caption{Testing for the null effect of functional coeﬃcient in the simulated data for scenario \textit{(S1)} under balanced cases \textit{(b)}. Empirical type-I error rates (top section) are reported for diﬀerent signiﬁcance level $\alpha=\{0.01, 0.05, 0.10, 0.15, 0.20, 0.30\}$ under $H_0$. Empirical rejection rates of the test (bottom section) are reported for $\rho =\{0.01, 0.10, 0.15, 0.20, 0.50, 1.00\}$ under $H_A$. Standard errors are reported in parenthesis.}
\label{test_table}
\noindent\makebox[\textwidth]{ 
\begin{tabular}{ccccccc}
 & \multicolumn{6}{c}{Size of $\mathcal{T}$} \\ \cline{2-7} 
$\sigma$ & $\alpha = 0.01$ & $\alpha = 0.05$ & $\alpha = 0.10$ & $\alpha = 0.15$ & $\alpha = 0.20$ & $\alpha = 0.30$ \\ \cline{2-7} 
$\text{1e-8}$ & 0.006 {[}0.001{]} & 0.048 {[}0.004{]} & 0.095 {[}0.005{]} & 0.145 {[}0.006{]} & 0.193 {[}0.007{]} & 0.308 {[}0.008{]} \\ \cline{2-7} 
$0.01$ & 0.001 {[}0.000{]} & 0.041 {[}0.002{]} & 0.094 {[}0.003{]} & 0.141 {[}0.004{]} & 0.187 {[}0.004{]} & 0.281 {[}0.005{]} \\ \cline{2-7} 
$0.10$ & 0.014 {[}0.001{]} & 0.056 {[}0.003{]} & 0.099 {[}0.003{]} & 0.152 {[}0.004{]} & 0.206 {[}0.005{]} & 0.293 {[}0.005{]} \\ \cline{2-7} 
 & \multicolumn{6}{c}{Power of $\mathcal{T}$ at $\alpha = 0.05$} \\ \cline{2-7} 
$\sigma$ & $\rho = 0.01$ & $\rho = 0.10$ & $\rho = 0.15$ & $\rho = 0.20$ & $\rho = 0.50$ & $\rho = 1.00$ \\ \cline{2-7} 
$\text{1e-8}$ & 0.026 {[}0.007{]} & 0.476 {[}0.022{]} & 0.992 {[}0.004{]} & 1.000 {[}0.000{]} & 1.000 {[}0.000{]} & 1.000 {[}0.000{]} \\ \cline{2-7} 
0.01 & 0.006 {[}0.003{]} & 0.488 {[}0.022{]} & 0.966 {[}0.008{]} & 1.000 {[}0.000{]} & 1.000 {[}0.000{]} & 1.000 {[}0.000{]} \\ \cline{2-7} 
0.10 & 0.008 {[}0.004{]} & 0.414 {[}0.022{]} & 0.980 {[}0.006{]} & 1.000 {[}0.000{]} & 1.000 {[}0.000{]} & 1.000 {[}0.000{]} \\ \cline{2-7} 
\end{tabular}
}
\end{table}

\section{Discussion}\label{s:discuss}

In this paper, we propose a novel scalar-on-image regression by borrowing the ideas from functional data analysis where we view images as functional measurements. We approximate both functional covariate and coefficient using the same orthonormal data-driven basis functions and cast the model into a GLM framework. The methodology relies on the assumption that the leading eigenbasis functions of the functional predictor are most predictive of generalized response and that the latent predictor signals are relatively smooth.

The proposed idea is applicable to other response types that follow multinomial, Poisson, Gaussian, or any other distribution belonging to the Exponential family; in this case, we need to select the corresponding link function appropriately (e.g., log link for Poisson). The methodology can be extended to the case with multiple functional predictors observed with or without noise and defined on diverse sampling designs. Numerical results using chest X-rays show excellent classification performances in detecting (a) pneumonia and healthy patients, (b) viral and bacterial pneumonia cases. The method performs competitively even when we have a smaller sample to train the model. Two numerical experiments based on the data application show the robustness of the methodology and exhibit efficacy in attaining the inferential properties in terms of size and power. Despite the increased flexibility, the method is computationally efficient (computation time is in seconds) and can be implemented with the existing freely available software.

This work also opens a few avenues for future exploration. One possibility is to extend this methodology to explore the association between scalar responses and images that are observed longitudinally in a sparse design. Such an idea is vital to study the prognosis of diseases. In this case, we need to borrow the concept of longitudinal data analysis and merge it into the proposed scalar-on-image regression.

\begin{figure}[ht]
	\centering
	\caption{Example of posteroanterior chest X-ray images for three adult COVID-19 patients.}
	\includegraphics[width=0.15\textwidth]{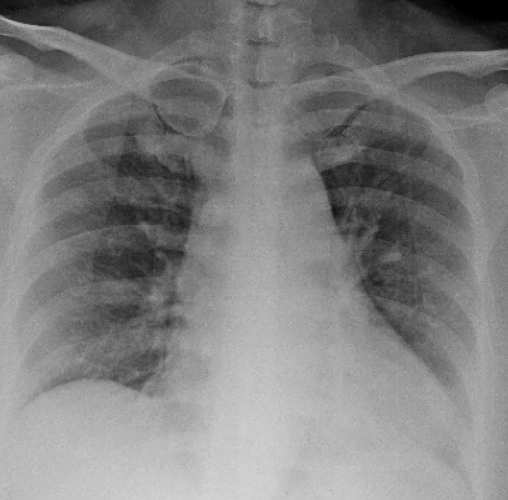} 
	\includegraphics[width=0.19\textwidth]{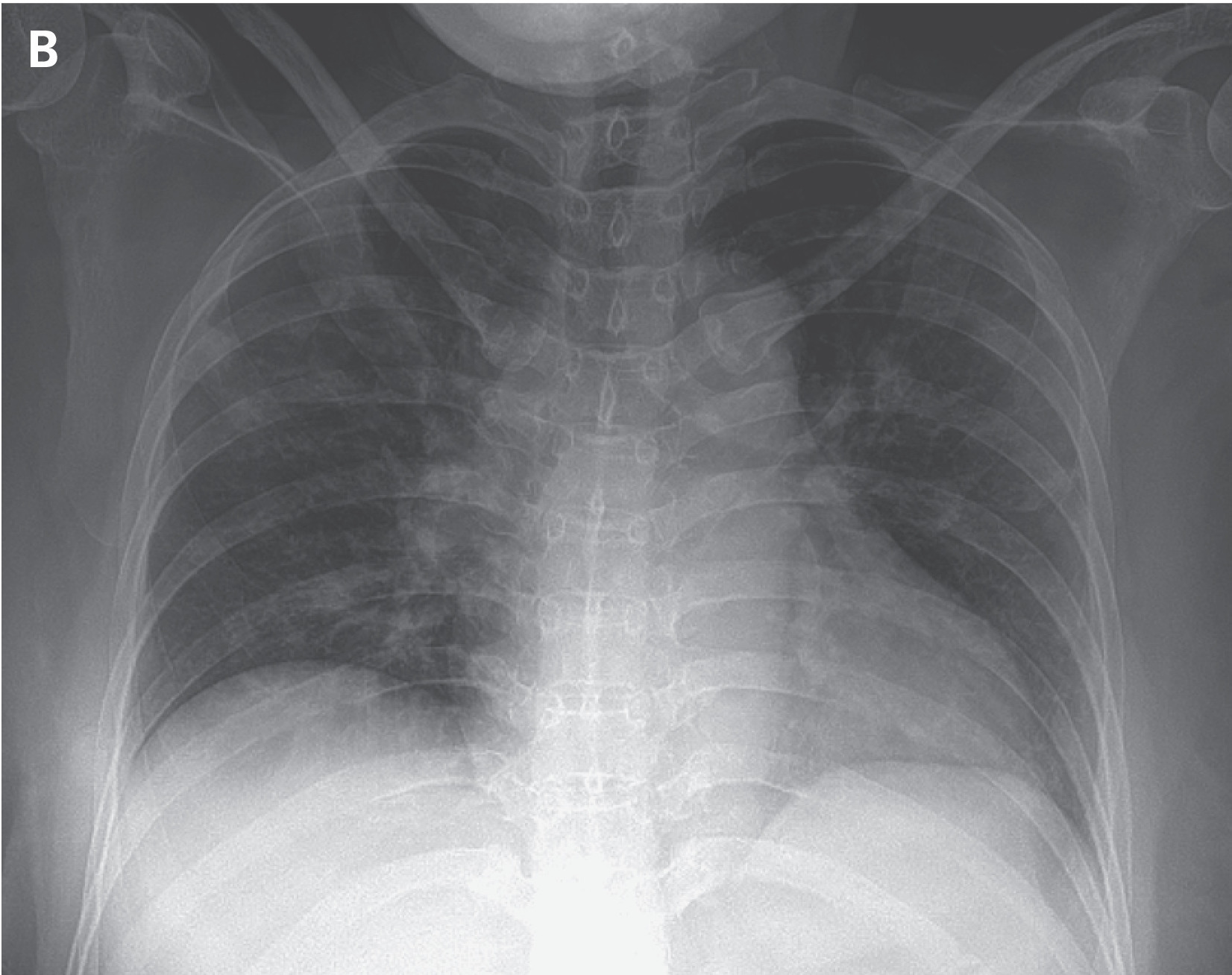} 
	\includegraphics[width=0.15\textwidth]{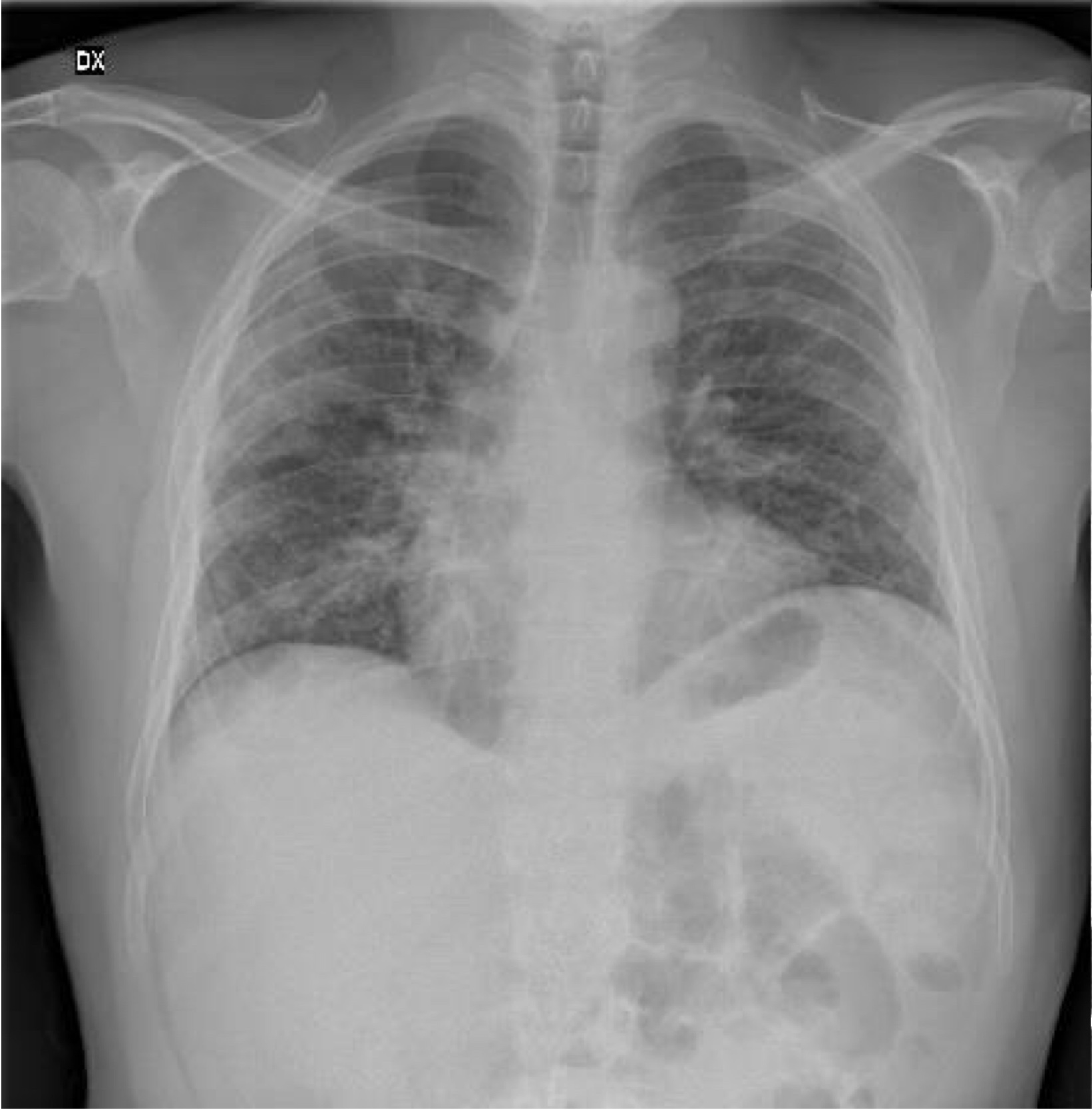} 
	\label{f:xray_COVID} 
\end{figure}

While the methodology is applied to the pediatric pneumonia cases, the approach can also be implemented in detecting pneumonia-like other infectious diseases. One possible test case could be the detection of COVID-19 patients in the recent pandemic. Based on the Johns Hopkins database \cite{jhu}, as of the 6th of May, 2020, approximately 3,755,341 people are infected by 2019 Novel coronavirus known as COVID-19 and more than 263,831 individuals died across the globe. Figure \ref{f:xray_COVID} illustrates the chest X-ray image for three COVID patients extracted from the freely available {\tt Kaggle} dataset \cite{cohen2020covid}; here the scattered white patches in both lungs are noticeable.
COVID-19 is very infectious, and once exposed, a person can transmit the infection being in an asymptomatic state, which is a state without showing medically known clinical signs or symptoms. Also, the detection of COVID-19 patients becomes a significant challenge due to the limited detection kits and short of medical supplies. The current testing approach is based on PCR, which is a time-consuming procedure. Besides, the community hospitals in remote areas are falling short in medical supplies and staff. Therefore, an alternative testing procedure based on radiography examination may play a role in identifying COVID patients as a supplement to the existing process. In a recent study, the chest radiography images of COVID-19 patients indicate abnormalities in their X-rays, suggesting the effectiveness of a radiography examination as an alternative infection detection approach; see \cite{huang2020clinical,ng2020imaging,ai2020correlation}. We want to address that COVID detection is more complicated due to the presence of pre-existing conditions and uncertainty about disease-biology as we are still in an early stage of this outbreak, and thus requires substantial validation and very high-quality data.




\section*{Acknowledgments}
The content is solely the responsibility of the author and does not represent the official views of the UnitedHealth Group R\&D or its any associated subsidiaries.


\bibliography{LDFR_TYPE_TBD}
\bibliographystyle{plain} 

\section{Supplementary Materials}\label{s:sup}
This Supplementary Material consists of two sections. Section~\ref{s:Image} provides additional results from the analysis of the image classification for healthy and pneumonia patients. Further results from the numerical experiment are illustrated in Section~\ref{s:Simul}.

\subsection{Additional results for pnumonia detection analysis} \label{s:Image}
\subsubsection{Exploratory analysis}
\begin{figure}[ht]
	\centering
	\caption{Image data $X_{i}(\cdot, \cdot)$ for three randomly chosen patients - healthy (left), viral (middle), and bacterial (right) pneumonia, respectively. We highlight two intensity profiles (in red and blue color) recorded at two specific locations in the $h$-direction. The dashed black lines correspond to the smoothed profiles.}
	\includegraphics[width=0.31\textwidth]{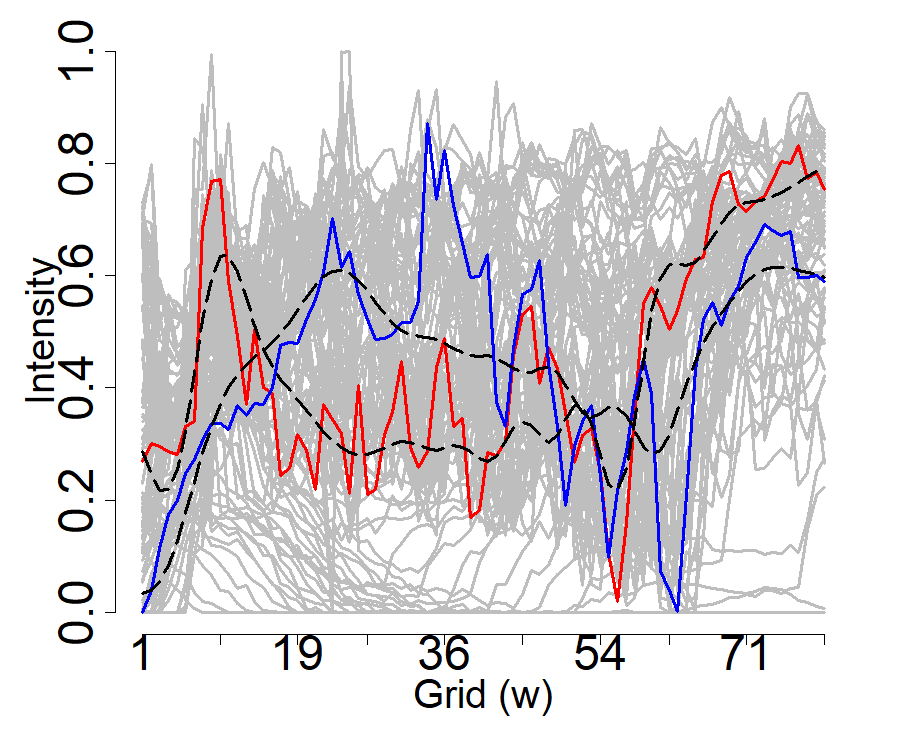}
      \includegraphics[width=0.31\textwidth]{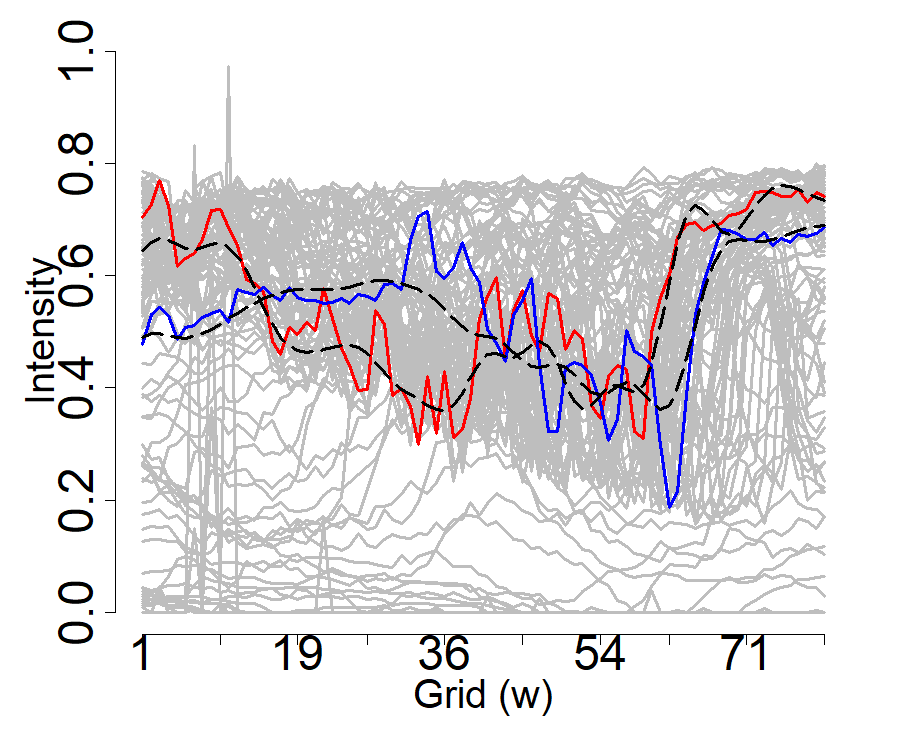}		
      \includegraphics[width=0.31\textwidth]{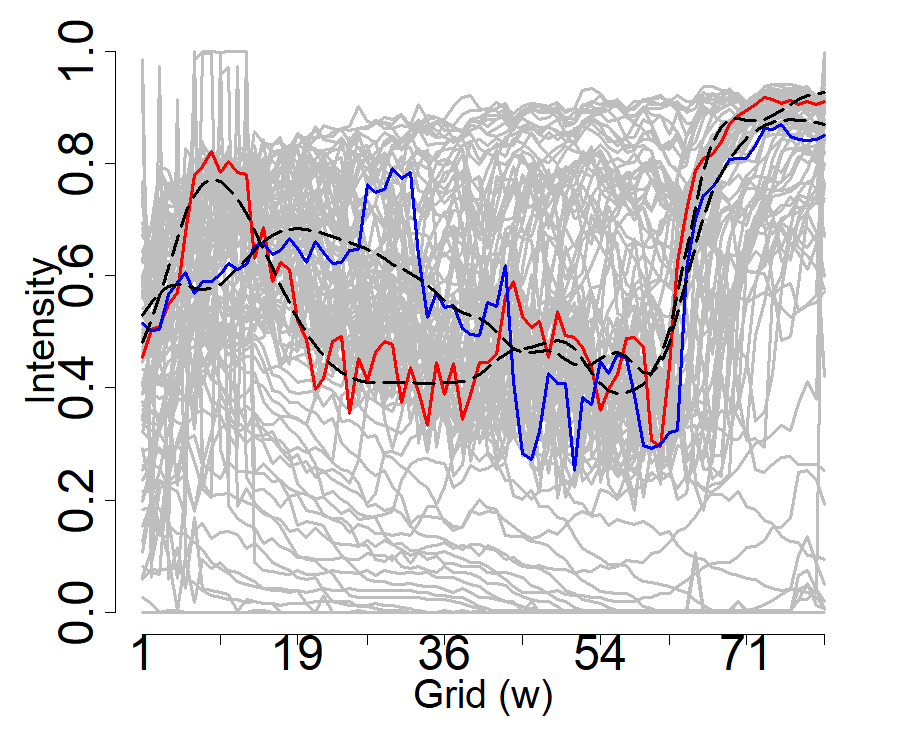}
	\label{f:xray_data} 
\end{figure}

Figure \ref{f:xray_data} illustrates gray-scaled image data for three randomly chosen patients - healthy (left), viral (middle), and bacterial (right) pneumonia cases, respectively. The observed functional covariate, $W_{i}(\cdot,\cdot),$  associated with the $i$th subject, is comprised of intensity profiles of the corresponding X-ray image containing values between 0 (black) and 1 (white). The corresponding smoothed profiles $\widehat W_{i}(\cdot, \cdot)$ are displayed in dashed black lines.
%
%
%

\subsubsection{Data analysis}
\begin{figure}[ht]
	\centering
	\caption{Top six estimated FPCs $\widehat \phi_{k}(\cdot).$ Corresponding  $\widehat \lambda_{k}$'s are reported in percentage.}
	\includegraphics[width=0.4\textwidth]{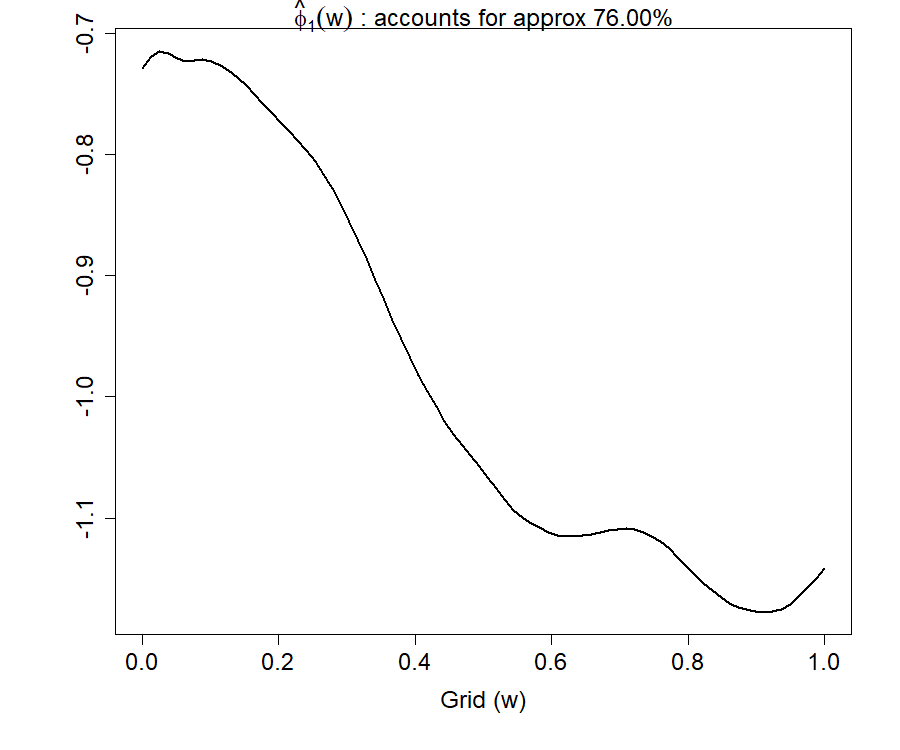}
      \includegraphics[width=0.4\textwidth]{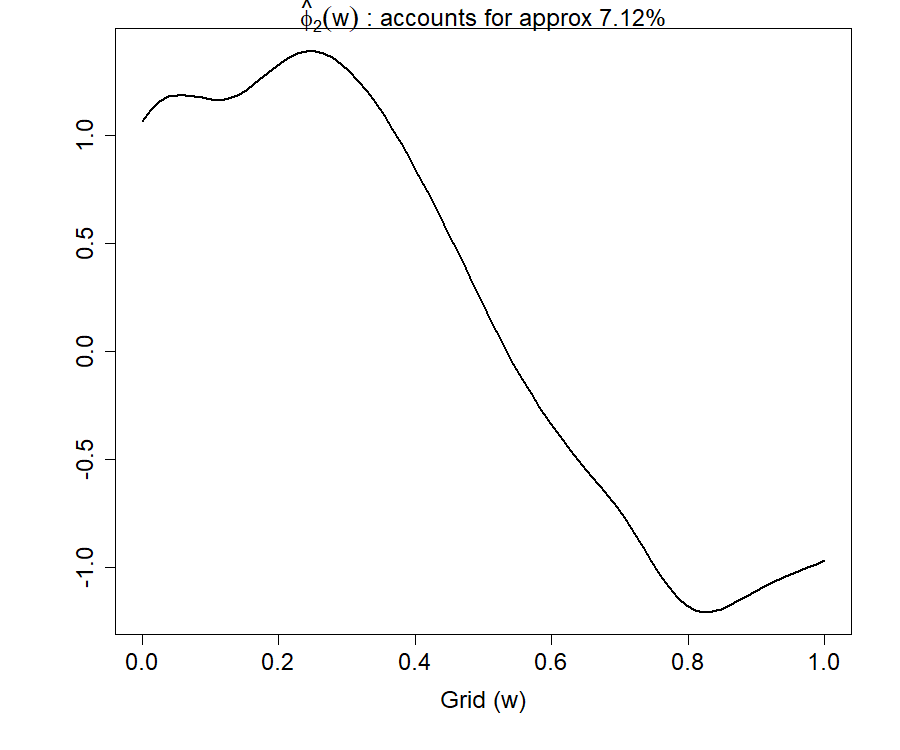}\\		
      \includegraphics[width=0.4\textwidth]{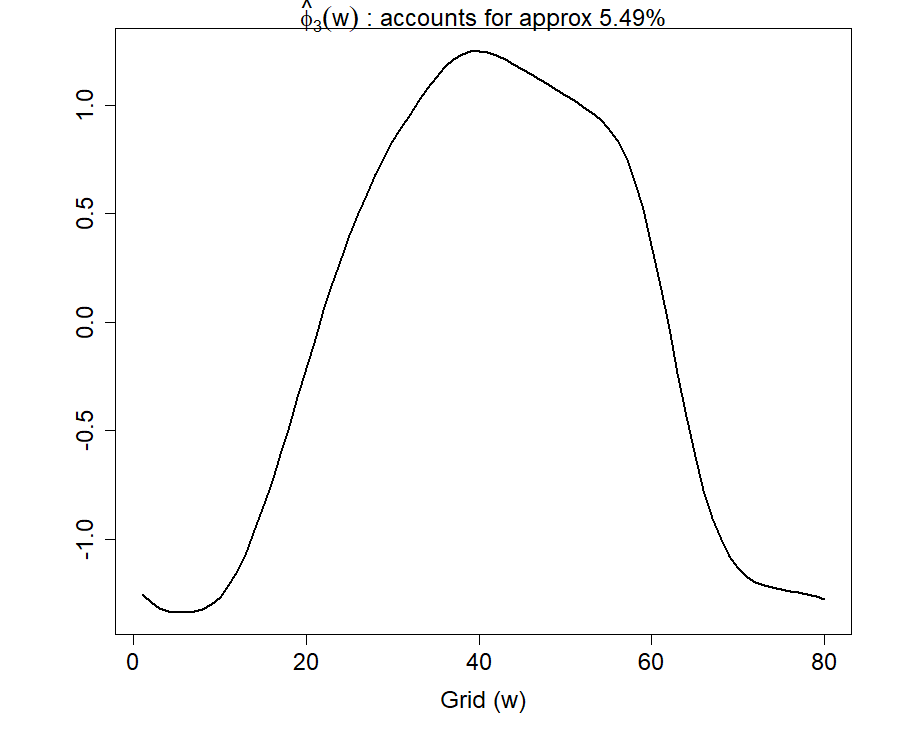}
      \includegraphics[width=0.4\textwidth]{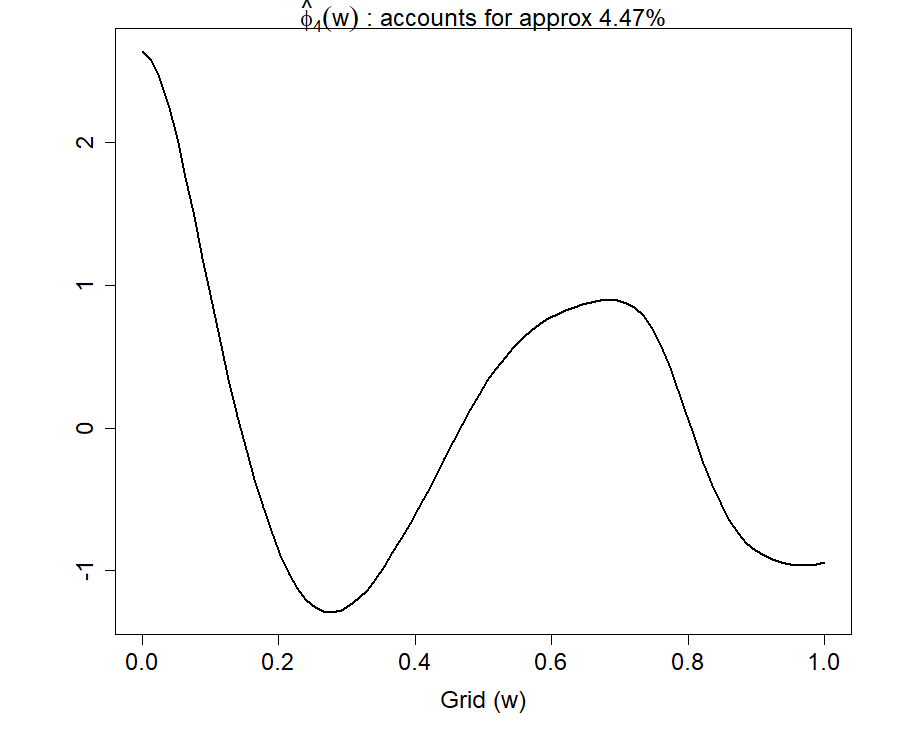}\\
      \includegraphics[width=0.4\textwidth]{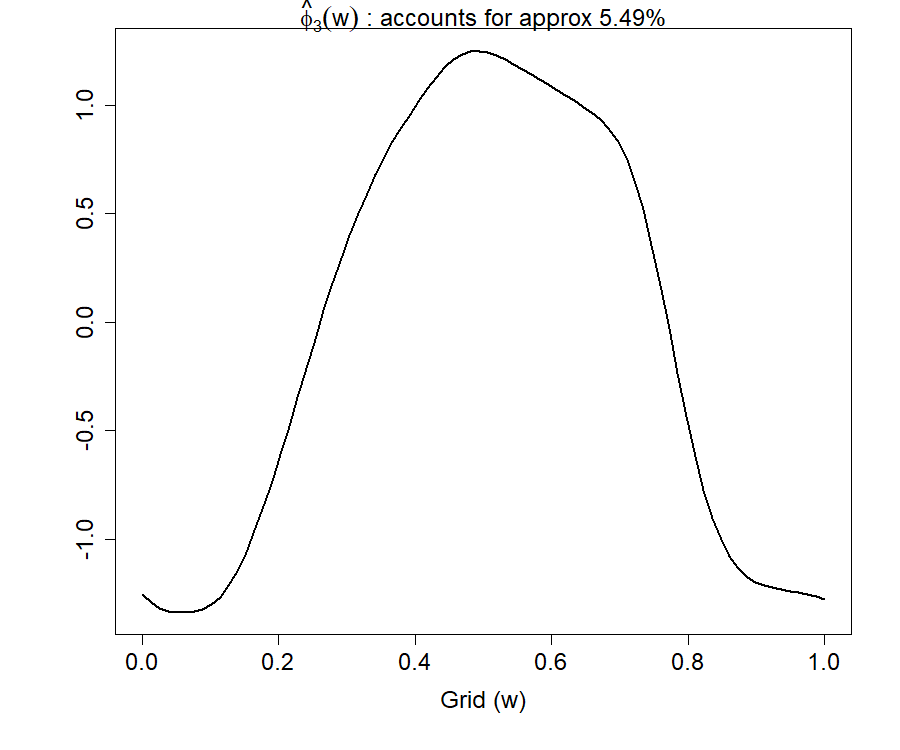}		
      \includegraphics[width=0.4\textwidth]{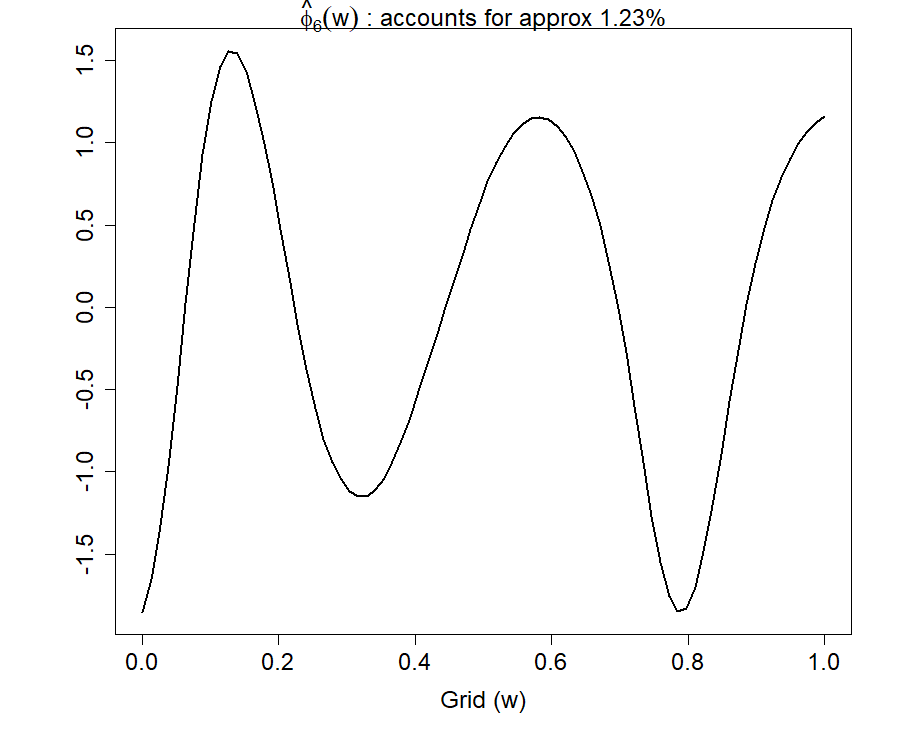}
	\label{f:xray_phi}  
\end{figure}

\begin{figure}[ht]
	\centering
	\caption{Top six estimated FPCs $\widehat \psi_{1l}(\cdot)$ associated with $\widetilde \xi_{w, ik}(\cdot)$ where $k = 1.$ Corresponding  $\widehat \eta_{kl}$'s are reported in percentage.}
	\includegraphics[width=0.4\textwidth]{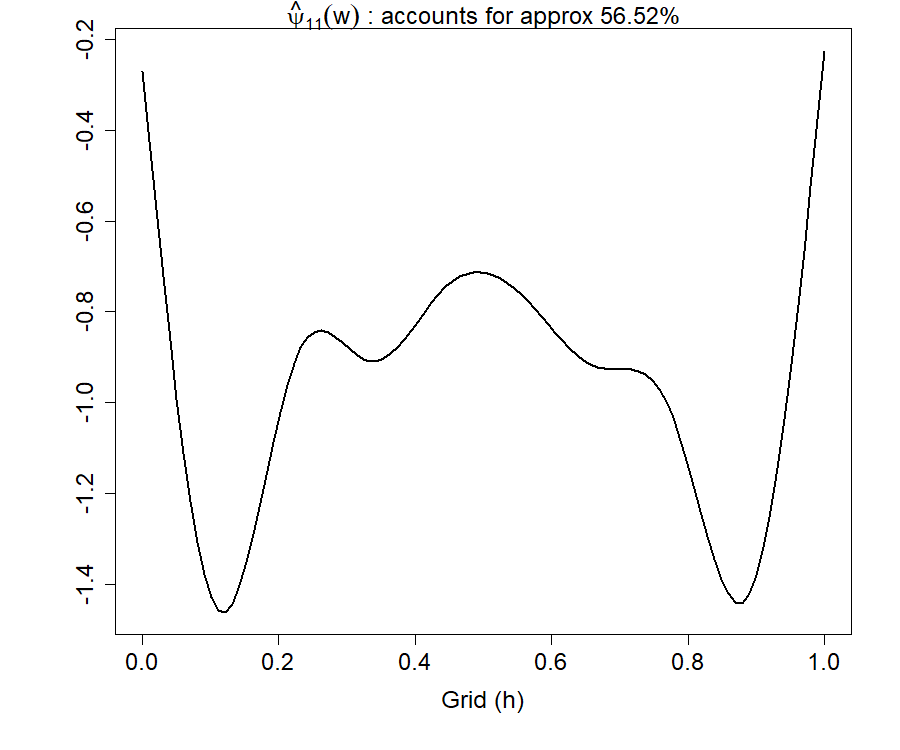}
      \includegraphics[width=0.4\textwidth]{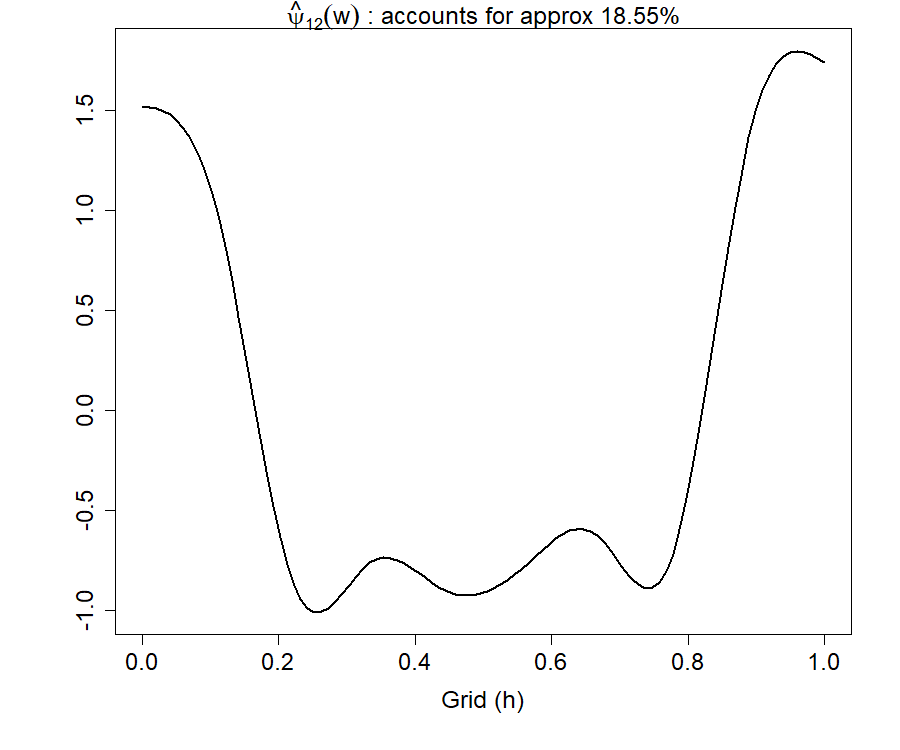}\\		
      \includegraphics[width=0.4\textwidth]{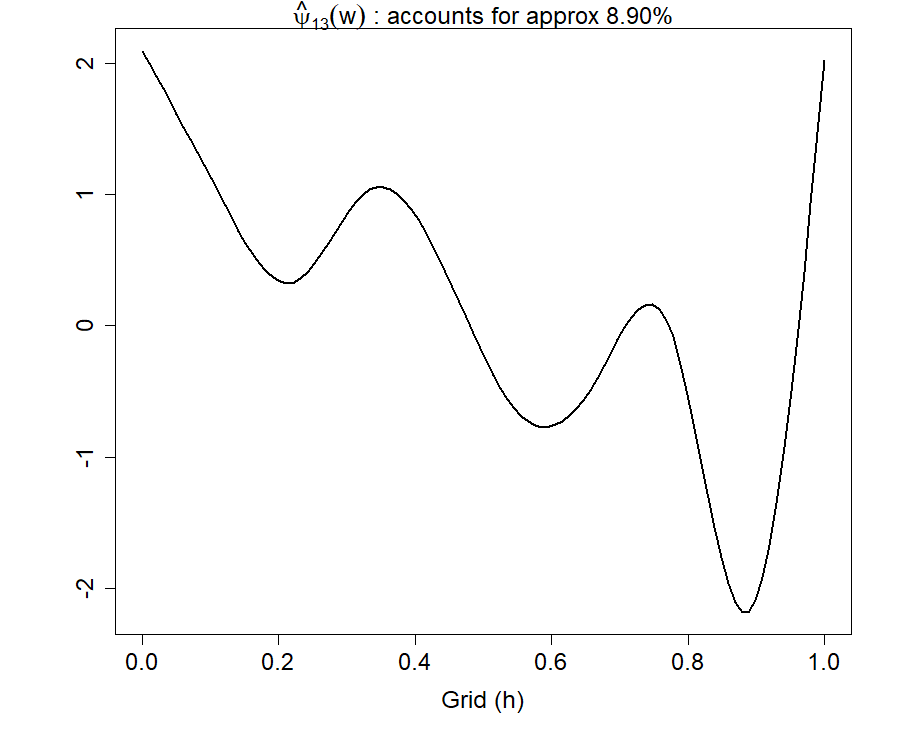}
      \includegraphics[width=0.4\textwidth]{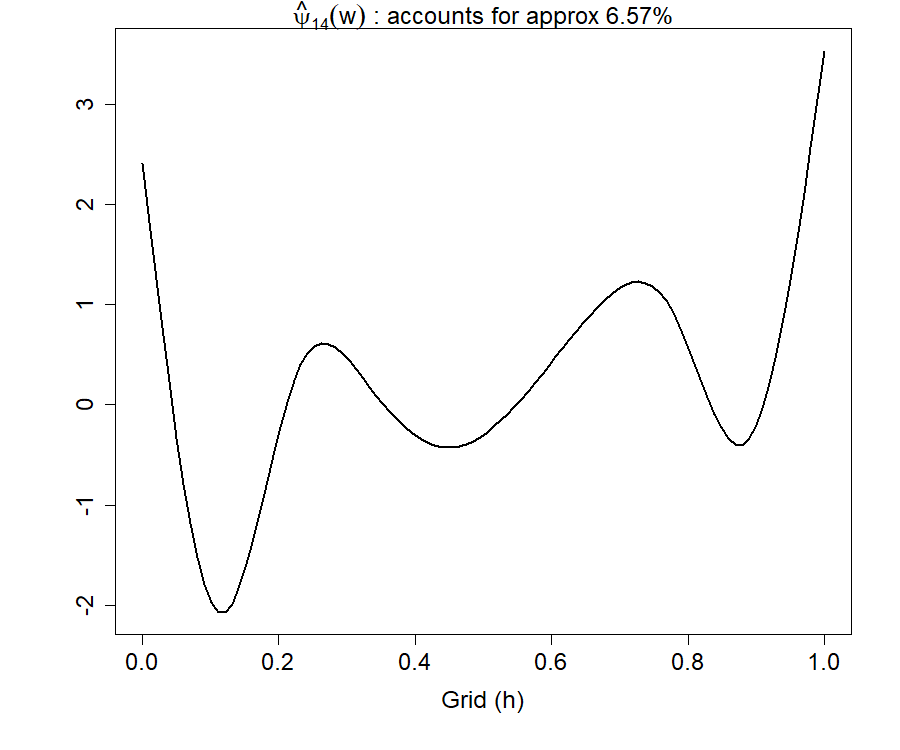}\\
      \includegraphics[width=0.4\textwidth]{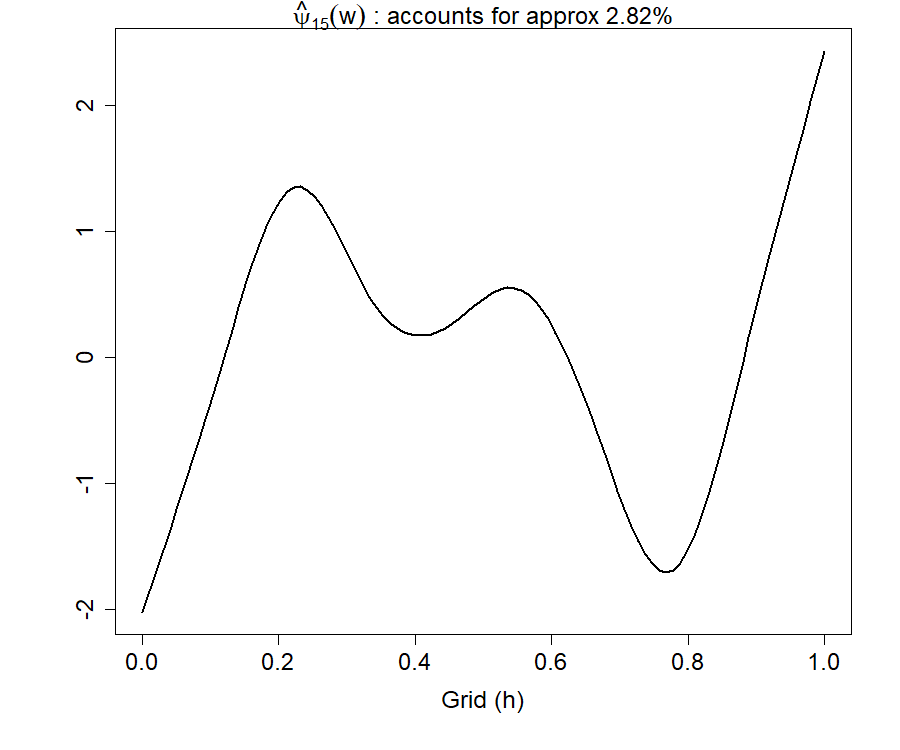}		
      \includegraphics[width=0.4\textwidth]{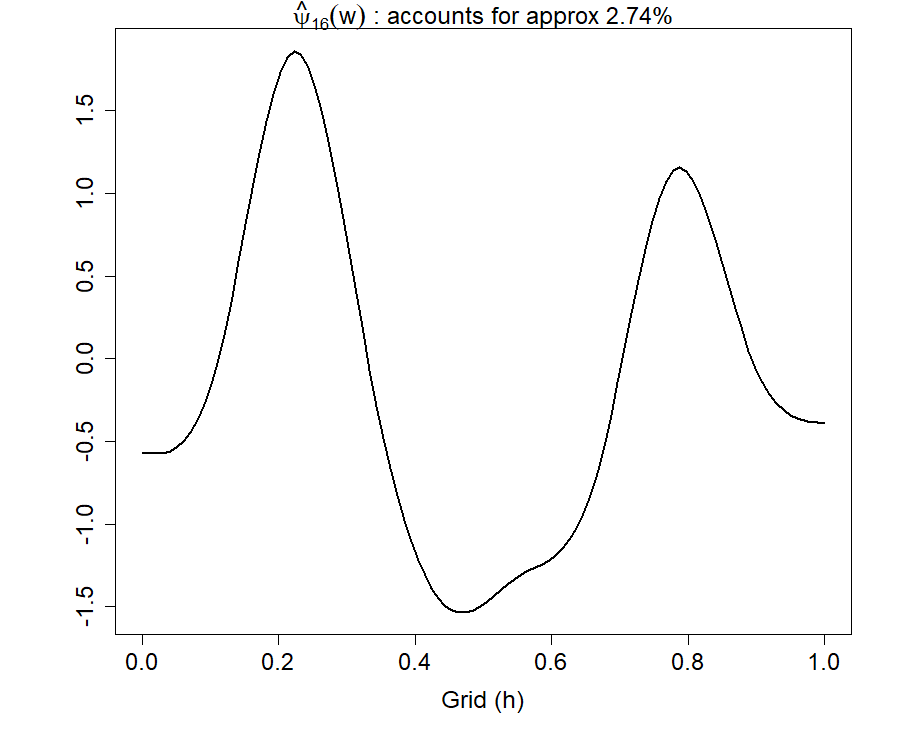}
	\label{f:xray_psi} 
\end{figure}

We illustrate the idea for \textit{(S1)}. We apply functional principal component (FPC) analysis on the marginal covariance function $\Gamma(w, w')$  induced by the functional predictor comprised of intensity profiles associated with 5,216 X-ray images. Setting a percentage of variance explained (PVE) value equal to 99\% provides directions at which we have maximum variation in the data. We estimate functional principal components $\widehat \phi_{k}(\cdot)$  and corresponding eigenvalues $\widehat \lambda_{k};$ where $k = 1, \cdots, 19$. Figure \ref{f:xray_phi} displays the estimated directions with the explained percentage of variance; first, 6 FPCs explain almost 96\% variation in the data. Similarly, spectral decomposition on $\widehat G_{k}(h, h')$ for each $k,$ results in $\widehat \psi_{kl}(\cdot)$'s and $\widehat \eta_{kl}$'s. Figure   \ref{f:xray_psi} displays the estimated directions  associated with the profile of loadings $\widetilde \xi_{w, i1}(\cdot)$; first, 6 FPCs explain approximately 96\% variation in the data. Similarly, we do it for each $k.$ In our example, $L_1 + \cdots + L_{19} = 187.$

\begin{figure}[ht]
	\centering
	\caption{Box plot of estimated scores $\widehat \zeta_{ikl}$'s for healthy and pneumonia patients. First, 60 loadings are displayed.}
\label{f:score_plot}
\noindent\makebox[\textwidth]{ 
	\includegraphics[width=1.1\textwidth]{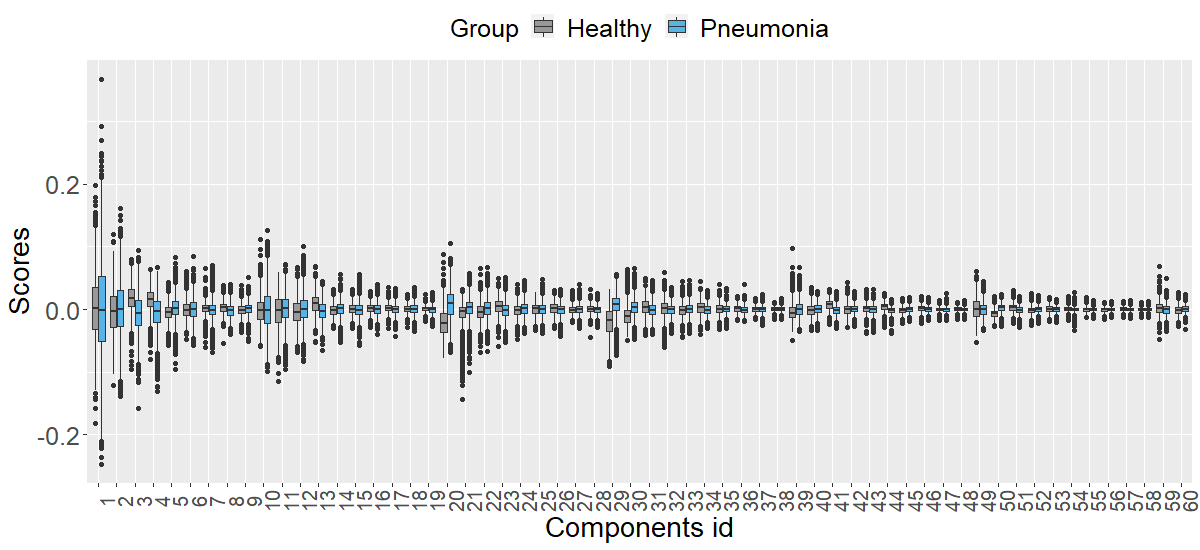}      
}
\end{figure}

\begin{figure}[ht]
	\centering
	\caption{Testing the equality of mean scores between healthy and pneumonia patients. Reported are the p-values in log scale; color codes are used to represent different $k = 1, \cdots, 19$. Reference line (black) is drawn representing the nominal type-I error rate after adjusting for multiple comparisons assuming $L_1 = \cdots = L_{17} \approx 10$.}
\label{f:score_pv}
\noindent\makebox[\textwidth]{ 
	\includegraphics[width=0.5\textwidth]{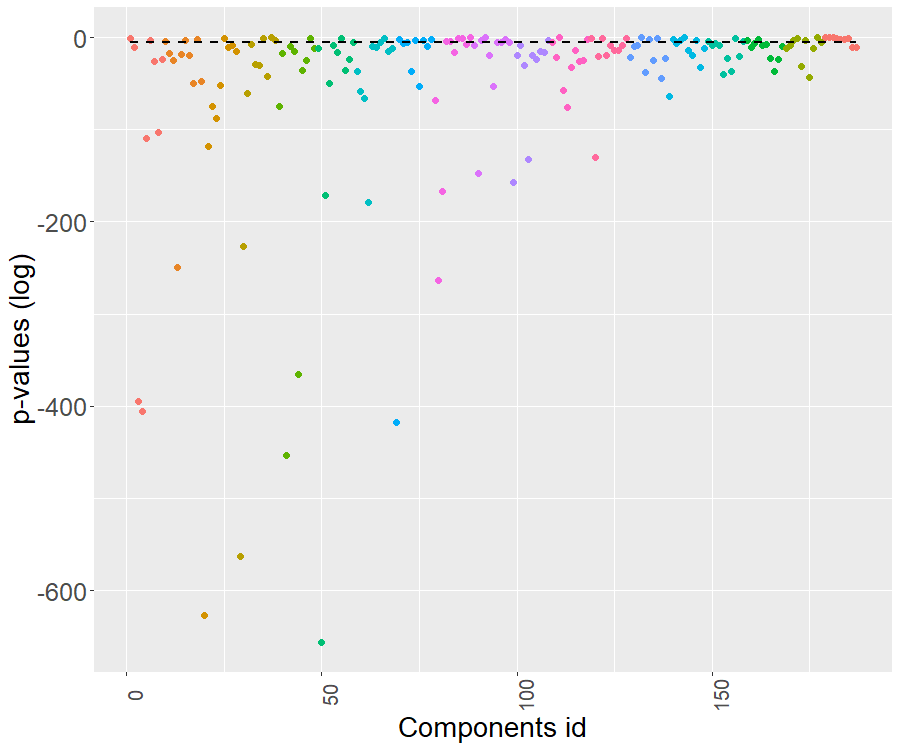}      
}
\end{figure}

Figure \ref{f:score_plot} displays the boxplots of the estimated scores (first 60 components) corresponding to both normal and pneumonia patients' intensity profiles. Figure \ref{f:score_pv} displays the log p-values for testing the equality of mean scores between healthy and pneumonia patients at 5\% level of significance. In particular, we test $H_{0}: {\bar \zeta}^{0}_{kl} =  {\bar \zeta}^{1}_{kl}$ for each $k$ and $l;$ where $\bar \zeta^{0}_{kl}$ and $\bar \zeta^{1}_{kl}$ refer to the mean values associated with the $kl$th component for normal and pneumonia patients, respectively. A two-tailed t-test statistic is used to make inference. For testing, $H_{0}: {\bar \zeta}^{0}_{k} = {\bar \zeta}^{1}_{k},$ for each $k,$ we adjust for multiple comparisons by Bonferroni correction.
%
%
%
%
%
%
%
%
%
%
%
%
%

We report additional classification metrics associated with a confusion matrix. The metrics are defined as in Section 6. For completeness, we recall the necessary items here. 
\begin{itemize}
\item True negative rate (TNR) = TN / (TN + FP)
 \item Positive predictive value (PPV) = TP / (TP + FP)
\item Mathhews correlation coefficient (MCC) = $(\text{TP} \cdot \text{TN} - \text{FP} \cdot \text{FN}) / \\ \sqrt{ \big\{ (\text{TP} + \text{FP}) \cdot (\text{TP} + \text{FN}) \cdot (\text{TN} + \text{FP}) \cdot (\text{TN} + \text{FN}) \big\} }$
\item F$_1$ = $2 \cdot \text{TPR}  \cdot \text{PPV} / (\text{PPV} + \text{TPR})$
\end{itemize} 
Table \ref{class_metrics12} reports results for \textit{(S1)}. We observe high TNRs implying low false-positive rates (FPRs), and high PPVs indicating low false discovery rates (FDR). Such observations are dominant across all settings. As before, PEN outperforms Non-PEN in the setting \textit{(M1)}, and the magnitude of improvement is higher in evaluating out-of-sample cases than that of in-sample ones. These findings are in alignment with what we observe in Table 1 of the draft.  

Table \ref{class_metrics22} provides additional results for \textit{(S1)}. The results between PEN and Non-PEN are competitive.

\begin{table}[ht]
\tiny
\caption{Comparison of chest X-rays between pneumonia and normal \textit{(S1)}. Classification metrics are presented for in-sample and out-of-sample data with respect to training-testing schemes \textit{(a)} and \textit{(b)} based on \textit{(M1)} and \textit{(M2)}. Median values and the corresponding IQR (in parenthesis) over 200 iterations are reported for scheme \textit{(a)}.}
\label{class_metrics12}
\noindent\makebox[\textwidth]{ 
\begin{tabular}{ccccccc}
\multicolumn{1}{l}{} & \multicolumn{3}{c}{Non-PEN} & \multicolumn{3}{c}{PEN} \\ \cline{2-7} 
Type & TNR & PPV & F$_1$ & TNR & PPV & F$_1$ \\ \cline{2-7} 
M1 + (a) + In & 0.940 {[}0.011{]} & 0.983 {[}0.003{]} & 0.967 {[}0.004{]} & 0.967 {[}0.010{]} & 0.990 & 0.976 {[}0.004{]} \\ \cline{2-7} 
M1 + (a) + Out & 0.931 {[}0.017{]} & 0.931 {[}0.016{]} & 0.931 {[}0.011{]} & 0.943 {[}0.023{]} & 0.944 {[}0.020{]} & 0.949 {[}0.010{]} \\ \cline{2-7} 
M1 + (b) + In & 0.936 & 0.982 & 0.972 & 0.968 & 0.991 & 0.982 \\ \cline{2-7} 
M1 + (b) + Out & 0.911 & 0.914 & 0.927 & 0.914 & 0.919 & 0.944 \\ \cline{2-7} 
M2 + (a) + In & 0.962 {[}0.015{]} & 0.962 {[}0.013{]} & 0.957 {[}0.004{]} & 0.959 {[}0.012{]} & 0.959 {[}0.011{]} & 0.956 {[}0.004{]} \\ \cline{2-7} 
M2 + (a) + Out & 0.946 {[}0.017{]} & 0.946 {[}0.017{]} & 0.939 {[}0.012{]} & 0.946 {[}0.020{]} & 0.945 {[}0.018{]} & 0.941 {[}0.011{]} \\ \cline{2-7} 
M2 + (b) + In & 0.966 & 0.966 & 0.955 & 0.959 & 0.959 & 0.957 \\ \cline{2-7} 
M1 + (b) + Out & 0.943 & 0.943 & 0.943 & 0.934 & 0.935 & 0.940 \\ \cline{2-7} 
\end{tabular}
}
\end{table}

\begin{table}[ht]
\tiny
\caption{Comparison of chest X-rays between viral and bacterial pneumonia \textit{(S2)}. Classification metrics are presented for in-sample and out-of-sample data with respect to training-testing schemes \textit{(a)} and \textit{(b)} based on \textit{(M1)} and \textit{(M2)}. Median values and the corresponding IQR (in parenthesis) over 200 iterations are reported for scheme \textit{(a)}.}
\label{class_metrics22}
\noindent\makebox[\textwidth]{ 
\begin{tabular}{ccccccc}
 & \multicolumn{3}{c}{Non-PEN} & \multicolumn{3}{c}{PEN} \\ \cline{2-7} 
Settings & TNR & PPV & F$_1$ & TNR & PPV & F$_1$ \\ \cline{2-7} 
M1 + (a) + In & 0.756 {[}0.052{]} & 0.866 {[}0.016{]} & 0.783 {[}0.027{]} & 0.715 {[}0.044{]} & 0.850 {[}0.013{]} & 0.791 {[}0.016{]} \\ \cline{2-7} 
M1 + (a) + Out & 0.711 {[}0.066{]} & 0.699 {[}0.036{]} & 0.685 {[}0.028{]} & 0.683 {[}0.051{]} & 0.693 {[}0.027{]} & 0.710 {[}0.025{]} \\ \cline{2-7} 
M1 + (b) + In & 0.749 & 0.864 & 0.793 & 0.756 & 0.864 & 0.778 \\ \cline{2-7} 
M1 + (b) + Out & 0.677 & 0.684 & 0.692 & 0.686 & 0.688 & 0.690 \\ \cline{2-7} 
M2 + (a) + In & 0.763 {[}0.046{]} & 0.760 {[}0.023{]} & 0.752 {[}0.014{]} & 0.693 {[}0.064{]} & 0.710 {[}0.027{]} & 0.729 {[}0.015{]} \\ \cline{2-7} 
M2 + (a) + Out & 0.695 {[}0.059{]} & 0.689 {[}0.030{]} & 0.685 {[}0.029{]} & 0.663 {[}0.066{]} & 0.682 {[}0.028{]} & 0.705 {[}0.030{]} \\ \cline{2-7} 
M2 + (b) + In & 0.783 & 0.773 & 0.755 & 0.680 & 0.708 & 0.741 \\ \cline{2-7} 
M2 + (b) + In & 0.711 & 0.700 & 0.687 & 0.666 & 0.685 & 0.705 \\ \cline{2-7} 
\end{tabular}
}
\end{table}

\subsection{Additional results for numerical experiment} \label{s:Simul}
Table \ref{simul_table2} displays additional binary classification results for the simulated data with different magnitude of noises. We notice similar phenomena as in Table 3 of the draft; the classification results for PEN and Non-PEN are very close. We observe high MCC values for smaller noise levels for $\sigma \leq 0.20$; where MCC is viewed as a correlation coefficient between observed and predicted binary responses. 

\begin{table}[ht]
\tiny
\caption{Comparison of chest X-rays between simulated pneumonia and normal \textit{(S1)} for balanced cases \textit{(b)}. Classification metrics (PPV, MCC, F$_1$) for in-sample and out-of-sample data at $\sigma \in \{0.01, 0.10, 0.20, 0.50, 1.00\}$ are summarized for Non-PEN and PEN estimation approach. Median values and the corresponding IQR (in parenthesis) are reported. Results are based on 500 simulations.}
\label{simul_table2}
\noindent\makebox[\textwidth]{ 
\begin{tabular}{cccccccc}
\multicolumn{1}{l}{} & \multicolumn{1}{l}{} & \multicolumn{3}{c}{Non-PEN} & \multicolumn{3}{c}{PEN} \\ \cline{3-8} 
$\sigma$ & Validation & PPV & MCC & F$_1$ & PPV & MCC & F$_1$ \\ \cline{3-8} 
$0.01$ & In & 0.945 {[}0.025{]} & 0.867 {[}0.023{]} & 0.931 {[}0.013{]} & 0.944 {[}0.027{]} & 0.862 {[}0.024{]} & 0.929 {[}0.014{]} \\ \cline{3-8} 
 & Out & 0.916 {[}0.047{]} & 0.804 {[}0.065{]} & 0.901 {[}0.032{]} & 0.919 {[}0.045{]} & 0.813 {[}0.065{]} & 0.901 {[}0.031{]} \\ \cline{3-8} 
$0.10$ & In & 0.941 {[}0.025{]} & 0.862 {[}0.022{]} & 0.929 {[}0.012{]} & 0.943 {[}0.027{]} & 0.858 {[}0.020{]} & 0.927 {[}0.012{]} \\ \cline{3-8} 
 & Out & 0.915 {[}0.053{]} & 0.803 {[}0.055{]} & 0.899 {[}0.033{]} & 0.918 {[}0.050{]} & 0.802 {[}0.058{]} & 0.898 {[}0.032{]} \\ \cline{3-8} 
$0.20$ & In & 0.908 {[}0.034{]} & 0.797 {[}0.025{]} & 0.897 {[}0.012{]} & 0.906 {[}0.029{]} & 0.796 {[}0.026{]} & 0.897 {[}0.013{]} \\ \cline{3-8} 
 & Out & 0.886 {[}0.055{]} & 0.747 {[}0.065{]} & 0.870 {[}0.034{]} & 0.884 {[}0.051{]} & 0.747 {[}0.066{]} & 0.871 {[}0.034{]} \\ \cline{3-8} 
$0.50$ & In & 0.813 {[}0.043{]} & 0.617 {[}0.036{]} & 0.806 {[}0.020{]} & 0.810 {[}0.043{]} & 0.620 {[}0.037{]} & 0.810 {[}0.021{]} \\ \cline{3-8} 
 & Out & 0.789 {[}0.060{]} & 0.573 {[}0.080{]} & 0.783 {[}0.043{]} & 0.787 {[}0.056{]} & 0.574 {[}0.082{]} & 0.784 {[}0.042{]} \\ \cline{3-8} 
$1.00$ & In & 0.788 {[}0.046{]} & 0.562 {[}0.035{]} & 0.778 {[}0.025{]} & 0.789 {[}0.050{]} & 0.562 {[}0.036{]} & 0.778 {[}0.024{]} \\ \cline{3-8} 
 & Out & 0.761 {[}0.063{]} & 0.507 {[}0.081{]} & 0.748 {[}0.048{]} & \multicolumn{1}{l}{0.763 {[}0.062{]}} & \multicolumn{1}{l}{0.508 {[}0.087{]}} & \multicolumn{1}{l}{0.747 {[}0.050{]}} \\ \cline{3-8} 
\end{tabular}
}
\end{table}

\end{document}